\documentclass[a4paper,12pt]{article}
\usepackage{amsfonts,amsthm,amsmath,amssymb,graphicx,hyperref,youngtab}

\newcommand{\be}{\begin{equation}}
\newcommand{\bea}{\begin{eqnarray}}
\newcommand{\ee}{\end{equation}}
\newcommand{\eea}{\end{eqnarray}}

\newcommand{\nn}{\nonumber}

\begin{document}

\makeatletter
\@addtoreset{equation}{section}
\makeatother
\renewcommand{\theequation}{\thesection.\arabic{equation}}

\rightline{WITS-CTP-122}
\vspace{1.8truecm}

\vspace{15pt}


{\LARGE{  
\centerline{\bf Higher Loop Nonplanar Anomalous Dimensions} 
\centerline{\bf from Symmetry} 
}}  

\vskip.5cm 

\thispagestyle{empty} \centerline{
    {\large \bf Robert de Mello Koch\footnote{ {\tt robert@neo.phys.wits.ac.za}}, Stuart Graham\footnote{\tt trautsgraham@gmail.com} 
                and Ilies Messamah\footnote{\tt Ilies.Messamah@wits.ac.za}} }

\vspace{.4cm}
\centerline{{\it National Institute for Theoretical Physics ,}}
\centerline{{\it Department of Physics and Centre for Theoretical Physics }}
\centerline{{\it University of Witwatersrand, Wits, 2050, } }
\centerline{{\it South Africa } }
\vspace{.4cm}

\vspace{1.4truecm}

\thispagestyle{empty}

\centerline{\bf ABSTRACT}

\vskip.4cm 

In this article we study the action of the one loop dilatation operator on operators with a classical dimension of order $N$.
These operators belong to the $su(2)$ sector and are constructed using two complex fields $Y$ and $Z$.
For these operators non-planar diagrams contribute already at the leading order in $N$ and the planar and large $N$ limits are distinct.
The action of the one loop and the two loop dilatation operator reduces to a set of decoupled oscillators and factorizes into an action on the
$Z$ fields and an action on the $Y$ fields.
Direct computation has shown that the action on the $Y$ fields is the same at one and two loops.
In this article, using the $su(2)$ symmetry algebra as well as structural features of field theory, we give compelling evidence
that the factor in the dilatation operator that acts on the $Y$s is given by the one loop expression, at any loop order.

\setcounter{page}{0}
\setcounter{tocdepth}{2}

\newpage

\tableofcontents

\setcounter{footnote}{0}

\linespread{1.1}
\parskip 4pt

{}~
{}~

\section{Summary and Conclusions}

The study of the spectrum of planar anomalous dimensions in ${\cal N}=4$ super Yang-Mills theory, motivated largely by the
AdS/CFT correspondence\cite{malda,Gubser:1998bc,Witten:1998qj}, has been extremely fruitful.
A key ingredient in this success has been the discovery that the planar dilatation operator can be identified with the Hamiltonian
of an integrable spin chain\cite{mz,bks,intreview}.
It is interesting to ask if integrability persists beyond the planar limit.
There has recently been some progress in this direction in \cite{Carlson:2011hy,Koch:2011hb,gs,deMelloKoch:2012ck}
which argues the action of the one loop dilatation operator on operators \cite{Balasubramanian:2004nb,dssi,dssii,bds,Bhattacharyya:2008rb}
with a classical dimension of order $N$, reduces to a set of decoupled oscillators.
The correlation functions of these operators receive contributions from non-planar diagrams
even at the leading order in a large $N$ expansion\cite{Balasubramanian:2001nh}.
This makes the study of correlators of these operators challenging.
Techniques employing group representation theory have been very effective for this 
problem\cite{cjr,dssi,dssii,Kimura:2007wy,bds,BHR1,Bhattacharyya:2008rb,BHR2,countconst,Kimura:2008ac,Kimura:2010tx,pasram,Kimura:2012hp}. 
There are two natural ways in which this work can be extended: one can try to study the action of the complete one
loop dilatation operator or one can try to extend known results to higher loops.
We will pursue the second goal in this article, considering the action of the dilatation operator at
higher loops in the $su(2)$ sector.
This sector is particularly simple, and so it provides the ideal setting in which to develop the necessary
methods.

The operators we study are built using the complex adjoint scalars $Z$ and $Y$.
We use $n$ $Z$ fields and $m$ $Y$ fields.
We will keep $n$ to be order $N$ and $m$ to be order $\sqrt{N}$.
The operators that we study, the restricted Schur polynomials, $O_{R,(r,s)\vec{\mu}}$ are labeled
by three Young diagrams $R$, $r$ and $s$ as well as a multiplicity label $\vec{\mu}$.
Young diagram $r$ contains $n$ boxes and the reader is encouraged to think of $r$ as a symmetric group
representation that organizes the $Z$ fields.
Similarly, $s$ is a Young diagram with $m$ boxes, and it together with the multiplicity label $\vec{\mu}$
organizes the $Y$ fields. 
Roughly speaking, the Young diagram $R$ (which has $m+n$ boxes) tells us how the two sets of fields are 
combined\footnote{Instead of saying that $R$ is a Young diagram with $m+n$ we will use the standard notation $R\vdash m+n$.
Similarly, $r\vdash n$ and $s\vdash m$.}.
The pair of Young diagrams $(r,s)$ label an irreducible representation of the $S_n\times S_m$ subgroup
of $S_{n+m}$.
Upon restricting to the $S_n\times S_m$ subgroup, the $S_{n+m}$ irreducible representation $R$ can subduce many copies of $(r,s)$.
The multiplicity labels $\vec{\mu}$ keep track of which copy of $(r,s)$ we are using. 
See equation (\ref{defrestschur}) below for the definition of the restricted Schur polynomial.
We study operators labeled by Young diagrams $r$ with two long rows.
These operators are dual to a system of two giant gravitons (built from the $Z$s) dressed by open strings
(given by the $Y$s). 

The one loop and two loop answers for the spectrum of anomalous dimensions show an interesting pattern.
The action of the dilatation operator at one loop and at two loops factorizes into a piece that acts only
on the $r$ label - i.e. on the $Z$ fields and a piece that acts only on the $s$ and $\vec{\mu}$ labels, i.e.
on the $Y$ fields.
Further at one loop and at two loop the factor that acts on the $Y$ fields is identical\cite{deMelloKoch:2012sv}.
This prompts a very natural question: does this persist at higher loops? In this article we will argue that
it does.

A brute force field theoretic approach to this problem seems hopeless.
Here however, we can take some guidance from progress made in the planar sector of the theory \cite{Beisert:2003ys}.
Indeed, working in the $su(2|3)$ sector of theory and using the symmetry algebra as well as structural features
from field theory, a great deal of information was obtained about higher loop corrections to the
dilatation operator\cite{Beisert:2003ys}. 
In the $su(2)$ sector that we study, we have operators $\vec{J}$ that generate an $SU(2)$ subgroup of the full
$SU(4)$ ${\cal R}$ symmetry enjoyed by the theory.
The $\vec{J}$ rotate the $Y$ and $Z$ fields amongst each other.
Since their eigenvalues are fixed by the $su(2)$ algebra, we know that these generators do not receive quantum corrections.
One of our results is a concrete expression for the action of these generators, in the large $N$ limit, 
on restricted Schur polynomials.
This is described in section 2.
In contrast to the operators $\vec{J}$ the dilatation operator does receive quantum corrections. 
Since the operators $\vec{J}$ commute with the dilatation operator, we do have some information about higher loop corrections.
Using this algebra, together with the large $N$ limit and the constraints that follow from the fact that the dilatation
operator is constructed by summing Feynman diagrams, we will give compelling evidence that the factor in the dilatation 
operator that acts on the $Y$s is given by the one loop expression at any loop order.
Concretely, the algebra $\big[\vec{J},D\big]=0$ implies a set of recursion relations, hermitticity of the dilatation
operator equates certain matrix elements of $D$ and the fact that we work at large $N$ implies that we can neglect
changes in Young diagram $r$ and further that the relation between $R$ and $r$ is preserved by 
$D$\footnote{$r$ is obtained by removing boxes from $R$. When we say that the relation between $R$ and $r$ is preserved by 
$D$, we mean that $D$ will only mix operators that are obtained by pulling the same number of boxes from each row of the 
big Young diagram $R$ to obtain $r$.}. 
The derivation of these recursion relations and the structure of the dilatation operator and a demonstration
that they determine the one loop dilatation operator is carried out in section 3.
This analysis is most easily extended to higher loops by employing a continuum limit.
The structure of this continuum limit is developed in section 4.
In section 5 we demonstrate that the recursion relations derived in section 3 are replaced by partial differential equations.
These partial differential equations describe all higher loops corrections to the dilatation operator.
As we explain in section 5, they can be solved rather completely.

The fact that the factor in the dilatation operator that acts on the $Y$s is given by the one loop expression at any loop order
is not completely unexpected.
Indeed, the diagonalization of this factor, achieved in general in \cite{deMelloKoch:2012ck}, gives the set of states that is
consistent with the Gauss Law constraints on a compact giant graviton world volume\cite{Balasubramanian:2004nb}.
We expect these constraints to be satisfied at any order in the loop expansion, because the Gauss Law is an exact statement.

For simplicity we have restricted ourselves to the sector of the theory that is dual to a system of two giant gravitons.
It would be straight forward but rather tedious to extend this to systems of more than two giant gravitons.
A much more interesting generalization is to go beyond the $su(2)$ sector, because symmetry is not very constraining in the $su(2)$ sector.
This follows because the dilatation operator is abelian and not part of a bigger algebra.
Restricted Schur polynomials for the $su(2|3)$ sector have been derived in \cite{pnm} and the use of symmetry
in this sector would represent a very interesting generalization.

Another problem that should be tackled is to determine the factor in the dilatation operator that acts on the $Z$ label.
Understanding this factor, together with the results of this paper, would allow a determination of the exact large $N$ anomalous
dimensions.
This is not as unexpected as one might expect.
Indeed, the operators we study are dual to giant gravitons.
One expects the local relativistic invariant world volume theory dynamics to emerge from the sector of the theory we are considering.
This picture suggests a relatively simple expression for the anomalous dimensions, determined by relativistic dispersion relations.
The simplicity we find in this paper is the first signal that this expectation is correct.
For closely related discussions see \cite{Berenstein:2013md,Berenstein:2013eya}.

\section{Action of $su(2)$ elements on restricted Schur polynomials}

In this section our goal is to compute the action of the generators $J_{\pm}$ and $J_3$ on restricted Schur polynomials. 
We will freely make use of the results obtained in \cite{Koch:2011hb} in this section.
Recall that in terms of the complex coordinates $z$ and $y$, we can realize the su(2) algebra as follows
\bea
   J_+=y{\partial\over\partial z},\quad    J_-=z{\partial\over\partial y},\quad   J_3=y{\partial\over\partial y}-z{\partial\over\partial z}\, .
\eea
This follows because $SU(2)$ rotates the complex coordinates into each other.
These generators close the usual algebra
\bea
   \big[J_+ ,J_-\big]=J_3,\qquad \big[J_3,J_\pm\big]=\pm 2 J_\pm\, .
   \label{su2algebra}
\eea
When acting on the restricted Schur polynomials the generators are
\bea
   J_+={\rm Tr}\left( Y{d\over dZ}\right),\quad
   J_-={\rm Tr}\left( Z{d\over dY}\right),\quad
   J_3={\rm Tr}\left( Y{d\over dY}\right)-{\rm Tr}\left( Z{d\over dZ}\right)\, .
\eea
This follows because the $SU(2)$ ${\cal R}$-symmetry rotates the matrices $Z$ and $Y$ into each other.
In what follows we will make use of the identity \cite{rajmike}
\bea
   {\rm Tr}(\sigma Y^{\otimes m}\otimes Z^{\otimes n})=\sum_{T,t,u,\vec{\nu}}{d_T n! m!\over d_t d_u (n+m)!}
      \chi_{T,(t,u)\vec{\nu}^*}(\sigma^{-1})
      \chi_{T,(t,u)\vec{\nu}}(Z,Y)
\eea
where if $\vec{\nu}=(\nu_1,\nu_2)$ then $\vec{\nu}^*=(\nu_2,\nu_1)$.
With a suitable choice of $\sigma$, the right hand side above gives any desired multitrace operator.
Thus, the above equation explains how to write an arbitrary multitrace operator as a linear combination of restricted Schur polynomials.
The sum above runs over all Young diagrams $T\vdash m+n$, $t\vdash n$ and $u\vdash m$ as well as over the multplicity labels $\vec{\nu}$.
$d_T$ denotes the dimension of the irreducible representation $T$ of $S_{n+m}$.
Similarly, $d_t$ denotes the dimension of irreducible representation $t$ of $S_{n}$ and $d_u$ the dimension of irreducible representation $u$ of $S_{m}$.
Finally, $\chi_{T,(t,u)\vec{\nu}^*}(\sigma^{-1})$ is the restricted character obtained by tracing $\Gamma_R(\sigma^{-1})$ over the
$(t,u)$ subspace.
The multiplicity index $\vec{\nu}^*=(\nu_2,\nu_1)$ tells us to trace the row index over the $\nu_2$ copy of $(r,s)$ and the column index over the
$\nu_1$ copy.

Consider a system of $g$ giant gravitons, i.e. the Young diagrams labeling the restricted Schur polynomials have a total of $g$ rows.
Our operators are built using $n$ $Z$ fields and $m$ $Y$ fields. 
Our operators are ($r\vdash n$ and $s\vdash m$)    
\bea
   \chi_{R,(r,s)\vec{\mu}}(Z,Y)= {1\over n! m!}\sum_{\sigma\in S_{n+m}}
   {\rm Tr}_{(r,s)\vec{\mu}}\left(\Gamma^R(\sigma )\right)
   {\rm Tr}(\sigma Y^{\otimes\, m}\,\otimes\, Z^{\otimes\, n})\, .
   \label{defrestschur}
\eea
In the above, $\vec{\mu}$ is a multiplicity label.
The restricted trace can be written in terms of an intertwining map $P_{R,(r,s)\vec{\mu}}$ as
\bea
   {\rm Tr}_{(r,s)\vec{\mu}}\left(\cdots\right)={\rm Tr}\left(P_{R,(r,s)\vec{\mu}}\cdots\right)
\eea
which factorizes as\cite{Koch:2011hb}
\bea
   P_{R,(r,s)\vec{\mu}}=p_{s\vec{\mu}}\otimes {\bf 1}_r
   \label{factorizeim}
\eea
It is possible to compute $P_{R,(r,s)\vec{\mu}}$ explicitely for restricted Schur polynomials that are labeled by Young diagrams 
$R$ with long rows and well separated corners\cite{Koch:2011hb}. 
We call this the displaced corners approximation.
Recall that $n\gg m$ and that $R$ has $g$ long rows.
We hold $g$ fixed and order 1 as we take $N\to\infty$.
In this limit the difference in the lengths of the corresponding rows of $R$ and $r$ can be neglected.
Let $V_g$ be a $g$ dimensional vector space.
In the construction of the projectors we removed $m$ boxes from $R$ to produce $r$ with each box represented by a vector in $V_g$.
The matrix $E_{ij}$ acting in $V_g$ is a $g\times g$ matrix with a 1 in the $i^{\rm th}$ row and $j^{\rm th}$ column, and zeros elsewhere.
The space $V_g^{\otimes \, k}$ obtained by tensoring $k$ copies of $V_g$ will also play a role in what follows.
The matrix $E^{(a)}_{ij}$ acts as $E_{ij}$ on the $a^{\rm th}$ copy of $V_g$ in $V_g^{\otimes \, k}$ and as the identity on all other copies.
In the displaced corners approximation the multiplicity label is a pair of Gelfand-Tsetlin patterns. 
Both the space $V_g^{\otimes \, k}$ as well as the $E^{(a)}_{ij}$ will play an important role in the computations that follow.
For more details and background see \cite{Koch:2011hb}.
Consider the action of $J_-$
\bea
   &&J_- \chi_{R,(r,s)\vec{\mu}}(Z,Y) = {\rm Tr}\left( Z{d\over dY}\right)\chi_{R,(r,s)\vec{\mu}}(Z,Y)\cr
   &&={m\over n! m!}\sum_{\sigma\in S_{n+m}}{\rm Tr}_{(r,s)\vec{\mu}}\left(\Gamma^R(\sigma )\right)
      {\rm Tr}(\sigma Y^{\otimes\, m-1}\,\otimes\, Z^{\otimes\, n+1})\cr
   &&={m\over n! m!}\sum_{\sigma\in S_{n+m}}{\rm Tr}_{(r,s)\vec{\mu}}\left(\Gamma^R(\sigma )\right)
      \sum_{T,(t^+,u^-)\vec{\nu}}{d_T (n+1)! (m-1)!\over d_{t^+} d_{u^-} (n+m)!}
      \chi_{T,(t^+,u^-)\vec{\nu}^*}(\sigma^{-1})
      \chi_{T,(t^+,u^-)\vec{\nu}}(Z,Y)\cr
   &&=\sum_{T,(t^+,u^-)\vec{\nu}}{d_T (n+1)\over d_{t^+} d_{u^-} (n+m)!}{(n+m)!\over d_T}\delta_{RT}
      {\rm Tr}_{R\oplus T}(P_{R,(r,s)\vec{\mu}}P_{T,(t^+,u^-)\vec{\nu}^*})
      \chi_{T,(t^+,u^-)\vec{\nu}}(Z,Y)\cr
   &&=\sum_{(t^+,u^-)\vec{\nu}}{n+1\over d_{t^+} d_{u^-} }
      {\rm Tr}_{R}(P_{R,(r,s)\vec{\mu}}P_{R,(t^+,u^-)\vec{\nu}^*})
      \chi_{R,(t^+,u^-)\vec{\nu}}(Z,Y)\, .
\eea
In the above expression $t^+$ is a Young diagram with $n+1$ boxes, $t^+\vdash n+1$.
The $+$ superscript indicates that a box has been added to $t$.
Similarly $u^-\vdash m-1$ with the $-$ superscript indicating that a box has been removed from $u$.
Let us now discuss how to perform the trace in the above expression. 
Using the factorized form of the intertwining map in (\ref{factorizeim}), we have\cite{Koch:2011hb}
\bea
  {\rm Tr}_{R}(P_{R,(r,s)\vec{\mu}}P_{R,(t^+,u^-)\vec{\nu}^*})
  ={\rm Tr}_{R}(p_{s\vec{\mu}}\otimes {\bf 1}_r \,\cdot\, p_{u^-\vec{\nu}^*}\otimes {\bf 1}_{t^+})\, .
\eea
The only way that this trace can be non-zero is if it is possible for $t^+$ to subduce $r$.
Write the projector ${\bf 1}_{t^+}$ in terms of its action on the m$^{\rm th}$ slot and ${\bf 1}_r$.
As an example to illustrate the idea, consider
\bea
   {\bf 1}_{\yng(3,2)}= E^{(m)}_{11}\, \otimes\, {\bf 1}_{\yng(2,2)} + E^{(m)}_{22}\, \otimes\, {\bf 1}_{\yng(3,1)}\, .
\eea
In the same way, if $t^{+\prime}_i=r$ we have\footnote{$t^{+\prime}_i$ is the Young diagram obtained by dropping a box from the i$^{\rm th}$ row of $t^+$.}
\bea
   {\bf 1}_{t^+}=E^{(m)}_{ii}\,\otimes\, {\bf 1}_r +\cdots
\eea
where $\cdots$ collects the terms that don't contribute to the value of the trace.
Consequently, in the displaced corners approximation we find\cite{Koch:2011hb}
\bea
{\rm Tr}_{R}(P_{R,(r,s)\vec{\mu}}P_{R,(t^+,u^-)\vec{\nu}^*})
  &=& {\rm Tr}_{R}(p_{s\vec{\mu}}\otimes {\bf 1}_r \,\cdot\, p_{u^-\vec{\nu}^*}\otimes {\bf 1}_{t^+})\cr
&=&\sum_i d_r {\rm Tr}_{V_g^{\otimes\, m}}(p_{s\vec{\mu}} \,\cdot\, p_{u^-\vec{\nu}^*}\otimes E^{(m)}_{ii})\delta_{t_i^{+\prime}r}\, .
\eea
To proceed further, recall that the multiplicity labels $\vec{\mu}$ and $\vec{\nu}$ stand for Gelfand-Tsetlin
patterns, that is, states of $U(g)$. 
In addition, $E_{ii}=|\vec{v}(i)\rangle\langle\vec{v}(i)|$ and there is no sum on $i$. 
The state $|\vec{v}(i)\rangle$ is a state in the fundamental of $U(g)$ - it is a $g$ dimensional vector of zeros except for the 
i$^{th}$ entry which is a 1.
The projector $p_{s\vec{\mu}}$ is\cite{Koch:2011hb}
\bea
   p_{s\vec{\mu}}=\sum_{a=1}^{d_s}|M_s^{\mu_1},a\rangle\langle M_s^{\mu_2},a|
\eea
where $|M_s^{\mu_1},a\rangle$ is a state labeled by a Gelfand-Tsetlin pattern.
$M_s^{\mu_1}$ is the pattern and $a$ labels states inside symmetric group irreducible representation $s$.
This state is obtained by taking a suitable linear combination of tensor products of $m$ copies (one for each slot) of the fundamental 
representation of $U(g)$. 
Rewrite this state as a linear combination of states which are each the tensor product of the fundamental representation for the $m^{\rm th}$ slot, 
with a state obtained by taking the tensor product of states of the remaining $m-1$ 
slots\footnote{It is useful to spell out the index structure of the next equation. The index $a$ runs over states
in $S_m$ irreducible representation $s$. The index $b$ runs over states in irreducible representations $s'$ subduced by $s$ when $S_m$ is 
restricted to $S_{m-1}$. We can thus put $a$ and the sets of different $b$ indices (one for every $s'$) into correspondence.}
\bea
  |M_s^{\mu_1},a\rangle = 
     \sum_{M_{s'}^{\alpha_1},M_F^l} C_{M_{s'}^{\alpha_1},M_F^l}^{M_s^{\mu_1}} 
        |M_{s'}^{\alpha_1},b\rangle\otimes |M_F^l\rangle\, .
\eea
$|M_F^l\rangle$ stands for a state in the fundamental representation of $U(g)$, $|M_F^l\rangle =|\vec{v}(l)\rangle$.
When $E^{(m)}_{ii}$ acts on $|M_s^{\mu_1},a\rangle$ it will pick out the piece with $l=i$. 
Thus,
\bea
  {\rm Tr}_{V_g^{\otimes\, m}}(p_{s\vec{\mu}} \,\cdot\, p_{u^-\vec{\nu}^*}\otimes E^{(m)}_{ii})
  &&=C_{M_{s'}^{\alpha_1},M_F^i}^{M_s^{\mu_1}} C_{M_{s'}^{\alpha_2},M_F^i}^{M_s^{\mu_2}}
    {\rm Tr}_{V_g^{\otimes\, m-1}}(p_{s' \vec{\alpha}} \,\cdot\, p_{u^-\vec{\nu}^*})\cr
  &&=d_{u^-}C_{M_{u^-}^{\nu_1},M_F^i}^{M_s^{\mu_1}} C_{M_{u^-}^{\nu_2},M_F^i}^{M_s^{\mu_2}}\, .
\eea
The Clebsch-Gordan coefficient can be written is in terms of bras and kets as follows
\bea
  C_{M_{u^-}^{\nu_1},M_F^i}^{M_s^{\mu_1}} = \langle \nu_1 \otimes \vec{v}(i) | \mu_1 \rangle\, .
\eea
Using this notation we finally have
\bea
{\rm Tr}_{R}(P_{R,(r,s)\vec{\mu}}P_{R,(t^+,u^-)\vec{\nu}^*})
=\sum_i d_r d_{u^-} \langle \mu_2 | \nu_2 \otimes \vec{v}(i) \rangle 
                  \langle \nu_1 \otimes \vec{v}(i) | \mu_1 \rangle\delta_{t_i^{+\prime}r}\, .
\eea
Thus,
\bea
   &&J_-\chi_{R,(r,s)\vec{\mu}}(Z,Y) =\sum_{(t^+,u^-)\vec{\nu}}
   {n+1\over d_{t^+} d_{u^-}} {\rm Tr}_{R}(P_{R,(r,s)\vec{\mu}}P_{R,(t^+,u^-)\vec{\nu}^*})\chi_{R,(t^+,u^-)\vec{\nu}}\cr
   &&=\sum_{(t^+,u^-)\vec{\nu}} \sum_i \delta_{RT} \delta_{t_i^{+\prime}r} {(n+1) d_r \over d_{t^+}}
   \langle \mu_2 | \nu_2 \otimes \vec{v}(i) \rangle \langle \nu_1 \otimes \vec{v}(i) | \mu_1 \rangle \chi_{R,(t^+,u^-)\vec{\nu}}\, .
\eea
We want the action on normalized operators.
The two point function of our operators are\cite{Bhattacharyya:2008rb}
\bea
  \langle \chi_{R,(r,s)\vec{\mu}}(Z,Y)\chi_{T,(t,u)\vec{\nu}}^\dagger (Z,Y)\rangle
  =\delta_{RT}\delta_{rt}\delta_{su}\delta_{\vec{\mu}\vec{\nu}}{f_R{\rm hooks}_R\over {\rm hooks}_r {\rm hooks}_s}\, .
\eea
By rescaling we can get operators with two point function equal to 1. Denote these by $O_{R,(r,s)\vec{\mu}}(Z,Y)$.
Acting on the normalized operators we have
\bea
   J_- O_{R,(r,s)\vec{\mu}}(Z,Y)=\sum_{T,(t^+,u^-)\vec{\nu}}
  (J_-)_{T,(t^+,u^-)\vec{\nu} \, ,\, R,(r,s)\vec{\mu}}O_{T,(t^+,u^-)\vec{\nu}}(Z,Y)
\eea
where
\bea
  &&(J_-)_{T,(t^+,u^-)\vec{\nu} \, ,\, R,(r,s)\vec{\mu}}
  =\sqrt{{\rm hooks}_r{\rm hooks}_s\over {\rm hooks}_{t^+}{\rm hooks}_{u^-}}\cr
  &&\times \sum_i \delta_{RT} \delta_{t_i^{+\prime}r} {(n+1) d_r \over d_{t^+}}
   \langle \mu_2 | \nu_2 \otimes \vec{v}(i) \rangle \langle \nu_1 \otimes \vec{v}(i) | \mu_1 \rangle\cr
  &&=\sqrt{{\rm hooks}_{t^+}{\rm hooks}_{s}\over {\rm hooks}_{r}{\rm hooks}_{u^-}}\sum_i \delta_{RT} \delta_{t_i^{+\prime}r} 
     \langle \mu_2 | \nu_2 \otimes \vec{v}(i) \rangle \langle \nu_1 \otimes \vec{v}(i) | \mu_1 \rangle\, .
  \label{Jminuselem}
\eea
Very similar arguments give
\bea
   J_+ O_{R,(r,s)\vec{\mu}}(Z,Y)=\sum_{T,(t^-,u^+)\vec{\nu}}
  (J_+)_{T,(t^-,u^+)\vec{\nu} \, ,\, R,(r,s)\vec{\mu}}O_{T,(t^-,u^+)\vec{\nu}}(Z,Y)
\eea
where
\bea
  &&(J_+)_{T,(t^-,u^+)\vec{\nu} \, ,\, R,(r,s)\vec{\mu}}
  =\sqrt{{\rm hooks}_r{\rm hooks}_s\over {\rm hooks}_{t^-}{\rm hooks}_{u^+}}\cr
  &&\times \sum_i \delta_{RT} \delta_{t^{-}r_i'} (m+1){d_s\over d_{u^+}}
    \langle \mu_2 \otimes \vec{v}(i) | \nu_2 \rangle \langle \nu_1 | \mu_1 \otimes \vec{v}(i) \rangle\cr
&& =\sqrt{{\rm hooks}_r{\rm hooks}_{u^+}\over {\rm hooks}_{t^-}{\rm hooks}_{s}} \sum_i \delta_{RT} \delta_{t^{-}r_i'}
    \langle \mu_2 \otimes \vec{v}(i) | \nu_2 \rangle \langle \nu_1 | \mu_1 \otimes \vec{v}(i) \rangle
   \label{Jpluselem}
\eea
and
\bea
   J_3 O_{R,(r,s)\vec{\mu}}(Z,Y)=\sum_{T,(t,u)\vec{\nu}}
  (J_3)_{T,(t,u)\vec{\nu} \, ,\, R,(r,s)\vec{\mu}}O_{T,(t,u)\vec{\nu}}(Z,Y)
\eea
where
\bea
  &&(J_3)_{T,(t,u)\vec{\nu} \, ,\, R,(r,s)\vec{\mu}}
  =\delta_{RT} \delta_{tr}\delta_{us}\delta_{\vec{\mu}\vec{\nu}} (m-n)\, .
  \label{J3element}
\eea

Our main interest is in the case of 2 rows.
This is the simplest setting in which to develop our arguments because in this case there are no multiplicities for the irreducible
representations that organize the $Y$ fields.
We will make use of a vector $\vec{m}$ which summarizes how to obtain $r$ from $R$.
Consider $O_{R,(r,s)}$. 
The vector $\vec{m}=(m_1,m_2)$ tells us how boxes should be removed from $R$ to obtain $r$.
Denoting the row lengths of $R$ by $(R_1,R_2)$ and of $r$ by $(r_1,r_2)$, we have $R_1=r_1+m_1$ and $R_2=r_2+m_2$.
As explained in Appendix E.1 of \cite{Koch:2011hb}, we can trade the irreducible representation $s$ organizing the $Y$ fields and $\vec{m}$ 
for an $SU(2)$ state.
In the new labelling, we specify an operator (which belongs to the sector of the theory constructed using $n$ $Z$s and $m$ $Y$s) by giving
the Young diagram $r$ and an $SU(2)$ state with labels $(j,j_3)$ where\footnote{$s_i$ denote the row lengths of $s$.}
\bea
   s=(j,j_3)\qquad \longleftrightarrow\qquad s_1={m+2j\over 2},\quad s_2={m-2j\over 2},\quad j_3={m_1-m_2\over 2}\, .
\eea
We will use the $j,j^3$ notation in what follows.

We know that $J_+$ removes a $Z$ box and adds a $Y$ box. 
Thus, it could have the following possible actions on $r$, the irreducible representation organizing the $Z$s
(the box to be removed has a $-$ sign in it - i.e. drop the box with the $-$ sign)
\bea
  \young(\,\,\,\,\,\,\,\,\,\,\,\,,\,\,\,\,\,\,\,)\longrightarrow &&  \young(\,\,\,\,\,\,\,\,\,\,\,-,\,\,\,\,\,\,\,)\cr
                                     j^3,r_1,r_2                 &&  j^3+{1\over 2},r_1-1,r_2\cr
\cr
&& {\rm OR}\cr
\cr
                                                 \longrightarrow &&  \young(\,\,\,\,\,\,\,\,\,\,\,\,,\,\,\,\,\,\,-)\cr
                                                                 &&  j^3-{1\over 2},r_1,r_2-1
\eea
It is trivial to understand how the row lengths $r_1$ and $r_2$ change when the box shown is dropped.
To understand the changes in $j^3$, note the following: $J_+$ does not change the shape of $R$ so that
if we know how $r$ changes, we know how $\vec{m}$ changes. 
In the first possibility above we remove a box from the first row of $r$ which implies that $m_1$ grows by 1 and hence 
that $j^3$ grows by ${1\over 2}$.
In the second possibility above we remove a box from the second row of $r$ which implies that $m_2$ grows by 1 and hence 
that $j^3$ decreases by ${1\over 2}$.
Since we have added a $Y$ box, $J_+$ can have the following action on $s$ (the box that has been addded has a $+$ in it)
\vfill\eject
\bea
  \young(\,\,\,\,\,\,\,\,\,\,\,\,,\,\,\,\,\,\,\,)\longrightarrow &&  \young(\,\,\,\,\,\,\,\,\,\,\,\,+,\,\,\,\,\,\,\,)\cr
                                     j  \qquad\qquad             &&  j+{1\over 2}\cr
\cr
&& {\rm OR}\cr
\cr
                                                 \longrightarrow &&  \young(\,\,\,\,\,\,\,\,\,\,\,\,,\,\,\,\,\,\,\,+)\cr
                                                                 &&  j-{1\over 2}
\eea
Consequently we have\footnote{Note that we don't need to display $r_2$ since $r_2=n-r_1$.}
\bea
  &&J_+O^{(n,m)}(r_1,j,j^3)=A_+O^{(n-1,m+1)}(r_1-1,j+{1\over 2},j^3+{1\over 2})
                           +B_+O^{(n-1,m+1)}(r_1-1,j-{1\over 2},j^3+{1\over 2})\cr
                         &&+C_+O^{(n-1,m+1)}(r_1,j+{1\over 2},j^3-{1\over 2})
                           +D_+O^{(n-1,m+1)}(r_1,j-{1\over 2},j^3-{1\over 2})\, .
\eea
We will describe the computation of $A_+$ in detail.
From (\ref{Jpluselem}) we have
\bea
   A_+ =\sqrt{{\rm hooks}_{r}\over {\rm hooks}_{t^-}}\sqrt{{\rm hooks}_{u^+}\over {\rm hooks}_{s}}
        \left(\langle j,j^3;{1\over 2},{1\over 2}|j+{1\over 2},j^3+{1\over 2}\rangle\right)^2
\eea
where
\bea
  &&\sqrt{{\rm hooks}_{r}\over {\rm hooks}_{t^-}}=\sqrt{(r_1+1)(r_1-r_2)\over (r_1-r_2+1)}\cr
  &&\sqrt{{\rm hooks}_{u^+}\over {\rm hooks}_{s}}=\sqrt{{m+2j+4\over 2}{2j+1\over 2j+2}}\cr
  &&\left(\langle j,j^3;{1\over 2},{1\over 2}|j+{1\over 2},j^3+{1\over 2}\rangle\right)^2={j+j^3+1\over 2j+1}\, .
\eea
Putting the above factors together, we find
\bea
  A_+=\sqrt{(r_1+1)(r_1-r_2)\over (r_1-r_2+1)}\sqrt{{m+2j+4\over 2}{2j+1\over 2j+2}}{j+j^3+1\over 2j+1}\, .
\eea
In the large $N$ limit this simplifies to
\bea
  A_+=\sqrt{r_1}\sqrt{{m+2j+4\over 2}{2j+1\over 2j+2}}{j+j^3+1\over 2j+1}\, .
\eea
Very similar arguments imply that
\bea
  && B_+=\sqrt{r_1}\sqrt{{m-2j+2\over 2}{2j+1\over 2j}}{j-j^3\over 2j+1}\, ,\cr
  && C_+=\sqrt{r_2}\sqrt{{m+2j+4\over 2}{2j+1\over 2j+2}}{j-j^3+1\over 2j+1}\, ,\cr
  && D_+=\sqrt{r_2}\sqrt{{m+2j+4\over 2}{2j+1\over 2j}}{j+j^3\over 2j+1}\, .
\eea

Next, consider the action of $J_-$.
We know that $J_-$ removes a $Y$ box and adds a $Z$ box. 
Thus, it could have the following possible actions on $r$, the irreducible representation organizing the $Z$s
(the box added has a $+$ sign in it)
\bea
  \young(\,\,\,\,\,\,\,\,\,\,\,\,,\,\,\,\,\,\,\,)\longrightarrow &&  \young(\,\,\,\,\,\,\,\,\,\,\,\,+,\,\,\,\,\,\,\,)\cr
                                     j^3,r_1,r_2                 &&  j^3-{1\over 2},r_1+1,r_2\cr
\cr
&& {\rm OR}\cr
\cr
                                                 \longrightarrow &&  \young(\,\,\,\,\,\,\,\,\,\,\,\,,\,\,\,\,\,\,\,+)\cr
                                                                 &&  j^3+{1\over 2},r_1,r_2+1
\eea
Since we have removed a $Y$ box, $J_-$ can have the following action on $s$ (the box removed has a $-$ in it)
\bea
  \young(\,\,\,\,\,\,\,\,\,\,\,\,,\,\,\,\,\,\,\,)\longrightarrow &&  \young(\,\,\,\,\,\,\,\,\,\,\,-,\,\,\,\,\,\,\,)\cr
                                     j  \qquad\qquad             &&  j-{1\over 2}\cr
\cr
&& {\rm OR}\cr
\cr
                                                 \longrightarrow &&  \young(\,\,\,\,\,\,\,\,\,\,\,\,,\,\,\,\,\,\,-)\cr
                                                                 &&  j+{1\over 2}
\eea
Consequently we have
\bea
  &&J_-O^{(n,m)}(r_1,j,j^3)=A_-O^{(n+1,m-1)}(r_1+1,j+{1\over 2},j^3-{1\over 2})
                           +B_-O^{(n+1,m-1)}(r_1+1,j-{1\over 2},j^3-{1\over 2})\cr
                         &&+C_-O^{(n+1,m-1)}(r_1,j+{1\over 2},j^3+{1\over 2})
                           +D_-O^{(n+1,m-1)}(r_1,j-{1\over 2},j^3+{1\over 2})\, .
\eea
To compute $A_-$, note that (\ref{Jminuselem}) implies that
\bea
   A_-=\sqrt{{\rm hooks}_{t^+}\over {\rm hooks}_{r}}\sqrt{{\rm hooks}_{s}\over {\rm hooks}_{u^-}}
       \left(\langle j,j^3|{1\over 2},{1\over 2};j+{1\over 2},j^3-{1\over 2}\rangle\right)^2
\eea
where
\bea
  &&\sqrt{{\rm hooks}_{t^+}\over {\rm hooks}_{r}}=\sqrt{(r_1+2)(r_1-r_2+1)\over (r_1-r_2+2)}\cr
  &&\sqrt{{\rm hooks}_{s}\over {\rm hooks}_{u^-}}=\sqrt{{m-2j\over 2}{2j+2\over 2j+1}}\cr
  &&\left(\langle j,j^3|{1\over 2},{1\over 2};j+{1\over 2},j^3-{1\over 2}\rangle\right)^2={j-j^3+1\over 2j+2}\, .
\eea
Thus, we find
\bea
  A_-=\sqrt{(r_1+2)(r_1-r_2+1)\over (r_1-r_2+2)}\sqrt{{m-2j\over 2}{2j+2\over 2j+1}}{j-j^3+1\over 2j+2}\, .
\eea
In the large $N$ limit this becomes
\bea
 A_-=\sqrt{r_1}\sqrt{{m-2j\over 2}{2j+2\over 2j+1}}{j-j^3+1\over 2j+2}\, .
\eea
Very similar arguments imply that
\bea
  && B_-=\sqrt{r_1}\sqrt{{m+2j+2\over 2}{2j\over 2j+1}}{j+j^3\over 2j}\, ,\cr
  && C_-=\sqrt{r_2}\sqrt{{m-2j\over 2}{2j+2\over 2j+1}}{j+j^3+1\over 2j+2}\, ,\cr
  && D_-=\sqrt{r_2}\sqrt{{m+2j+2\over 2}{2j\over 2j+1}}{j-j^3\over 2j}\, .
\eea
Using these results it is straight forward to find
\bea
  [J_+,J_-]O^{(n,m)}(r_1,j,j^3)=-nO^{(n,m)}(r_1,j,j^3)\, .
\eea
Noting that $J_3 O^{(n,m)}(r_1,j,j^3) =(m-n)O^{(n,m)}(r_1,j,j^3)$, this is indeed the correct large $N$ limit of (\ref{su2algebra}).

\section{Recursion relations and one loop dilatation operator}

The one loop dilatation operator in the $su(2)$ sector\cite{Beisert:2003tq}
\bea
   D_2= - g_{\rm YM}^2 {\rm Tr}\,\big[ Y,Z\big]\big[ \partial_Y ,\partial_Z\big]
   \label{oneloopD}
\eea
acting on two giant graviton systems, is given by \cite{Carlson:2011hy,Koch:2011hb}
\bea
   D_2 O^{(n,m)}(r_1,j,j^3) &&=
g_{YM}^2 \left[-{1\over 2}\left( m-{(m+2)(j^3)^2\over j(j+1)}\right)
\Delta O^{(n,m)}(r_1,j,j^3)\right.\nonumber \\
&& + \sqrt{(m + 2j + 4)(m - 2j)\over (2j + 1)(2j+3)} {(j+j^3 +
1)(j-j^3 + 1) \over 2(j + 1)}
 \Delta O^{(n,m)}(r_1,j+1,j^3)
\nonumber \\
&& \left. +\sqrt{(m + 2j + 2)(m - 2j +2)\over (2j + 1)(2j-1)}
{(j+j^3 )(j-j^3 ) \over 2 j} \Delta O^{(n,m)}(r_1,j-1,j^3)
\right]\nonumber\\
\label{basicdil}
\eea
where ($r_2=n-r_1$)
\bea
\Delta O^{(n,m)}(r_1,j,j^3) &&=\sqrt{(N+r_1)(N+r_2)}(O^{(n,m)}(r_1+1,j,j^3)+O^{(n,m)}(r_1-1,j,j^3))\nonumber \\
 &&-(2N+r_1+r_2)O^{(n,m)}(r_1,j,j^3)\, .\label{nicecombo}
\eea
Our goal in this section is to argue that we can recover (\ref{basicdil}) by requiring that the correct algebra
\bea
 \big[ D_2, J_\pm\big]=0=\big[D_2,J_3\big]
 \label{rightalgebra}
\eea
is obeyed. 
We have already obtained a formula for the action of $J_{\pm}$ and $J_3$ on restricted Schur polynomials.
Our first task is thus to obtain a similar result for the action of $D_2$, that can be used in (\ref{rightalgebra}).
We are not trying to write down a detailed formula for $D_2$, but rather, want to write the general structure of this
action that is consistent with the fact that it is derived by summing Feynman diagrams, we are working at large $N$ 
and the dilatation operator is a hermittian operator. 
Given this general form, we will derive the detailed matrix elements by requiring (\ref{rightalgebra}).

There is a pair of derivatives in the one loop dilatation operator (\ref{oneloopD}).
Since they share an index, their action on the restricted Schur polynomials produces a Kronecker delta function.
Equivalently, at one loop our Feynman diagrams have a single interaction vertex and this vertex has two pairs of adjacent
fields, $Z,Y$ and $Z^\dagger,Y^\dagger$. 
Wick contraction with the vertex will thus set a pair of indices equal, producing a Kronecker delta function.
The net consequence of this Kronecker delta function is that the sum over $S_{n+m}$ appearing in the evaluation of $D_2$
is reduced to a sum over the subgroup $S_{n+m-1}$\cite{DeComarmond:2010ie}. 
When we sum over the $S_{n+m-1}$ subgroup, the fundmental orthogonality relation forces one of the representations 
of $S_{n+m-1}$ subduced by $T$ to be equal to one of the representations subduced by $R$.
This allows $D_2$ to shift the position of a single box in each of the Young diagram labels of the restricted Schur polynomial.
At $p$-loops we will have $p$ insertions of the interaction vertex producing (at most) $p$ Kronecker delta functions, thereby
reducing the sum over $S_{n+m}$ to a sum over $S_{n+m-p}$.
This allows the $p$-loop dilatation operator to shift the position of (at most) $p$-boxes in each of the Young diagram labels 
of the restricted Schur polynomial.
Returning to one loop, a single box shifts position under the action of $D_2$. 
This implies that we can have the following changes in the labels of our operators
\bea
   j^3\to j^3,j^3\pm 1\,,\cr
   j  \to j,j\pm 1\,,\cr
   r_1\to r_1,r_1\pm 1\,.
\eea
This change of labels implies a total of 27 possible terms under the action of $D_2$
\bea
  D_2 O^{(n,m)}(r_1,j,j^3)=\sum_{c=-1}^1\sum_{d=-1}^1\sum_{e=-1}^1 \beta^{(n,m)}_{r_1,j,j^3}(c,d,e)O^{(n,m)}(r_1+c,j+d,j^3+e)
  \label{dilact}
\eea
This is slightly too general, as we have not yet put in the constraint that only 1 box can move, i.e. that even if $R$ and $T$ don't agree, 
by removing a single box from $R$ and a single box from $T$ we can get Young diagrams which do agree. 
The boxes that must be moved between $R$ and $T$ can be deduced from the boxes moving between $r$ and $t$ and the number of $Y$ boxes 
that move between the rows (determined by $j^3$). 
The matrix element of the dilatation operator that takes
\bea
   O^{(n,m)}_{r_1,j,j^3}\longrightarrow O^{(n,m)}_{r_1+a, j+b,j^3+c}\equiv O^{(n,m)}_{t_1, j',j^{3\prime}} 
\eea
is $\beta^{(n,m)}_{r_1,j,j^3}(a,b,c)$.
The integer $a$ determines how $r_1$ changes, $t_1-r_1=a$.
The integer $c$ determines how $j^3$ changes, $j^{3 \prime}-j^3=c$.
From the definition of $j^3$ we have
\bea
   2j^3          &=& (R_1-r_1)-(R_2-r_2)\,,\\
   2j^{3 \prime} &=& (T_1-t_1)-(T_2-t_2)\,.
\eea
We also know that $T_1+T_2=R_1+R_2=m+n$ and $t_1+t_2=r_1+r_2 = n$ so that
\bea
   2j^3          &=& 2 R_1-(m+n) - 2 r_1 + n\,,\\
   2j^{3 \prime} &=& 2 T_1-(m+n) - 2 t_1 + n\,.
\eea
Subtracting these last two equations gives
\bea
   2(j^{3 \prime}-j^3)= 2c = 2(T_1-R_1)+2(r_1-t_1) = 2(T_1-R_1) - 2a\, .
\eea
Thus, $T_1-R_1 = a+c$ and we must have $ |a+c|\le 1$. This forces
\bea
  \beta^{(n,m)}_{r_1,j,j^3}(1,0,1)&=&0\qquad \beta^{(n,m)}_{r_1,j,j^3}(-1,0,-1)=0\nn\\
  \beta^{(n,m)}_{r_1,j,j^3}(1,1,1)&=&0\qquad \beta^{(n,m)}_{r_1,j,j^3}(-1,1,-1)=0\nn\\
  \beta^{(n,m)}_{r_1,j,j^3}(1,-1,1)&=&0\qquad \beta^{(n,m)}_{r_1,j,j^3}(-1,-1,-1)=0\,.
\eea
This reduces the number of terms in the action of $D_2$ to 21.
Next, we know that the dilatation operator is hermittian $D_2=D_2^\dagger$. 
This implies that
\bea
   \langle r_1+a,j+b,j^3+c| D_2 |r_1,j,j^3\rangle   =   \langle r_1,j,j^3| D_2 |r_1+a,j+b,j^3+c\rangle\, .
\eea
Further, since
\bea
   \langle r_1+a,j+b,j^3+c| D_2 |r_1,j,j^3\rangle   =   \beta^{(n,m)}_{r_1,j,j^3}(a,b,c)
\eea
and
\bea
   \langle r_1,j,j^3| D_2 |r_1+a,j+b,j^3+c\rangle   =   \beta^{(n,m)}_{r_1+a,j+b,j^3+c}(-a,-b,-c)
\eea
we find that the condition $D_2=D_2^\dagger$ implies that
\bea
    \beta^{(n,m)}_{r_1,j,j^3}(a,b,c)     =     \beta^{(n,m)}_{r_1+a,j+b,j^3+c}(-a,-b,-c)\, .
\eea
This reduces the number of unknown terms to be determined to 11. 

In the large $N$ limit, the string coupling $g_s={1\over N}$ goes to zero.
Consequently there is no string splitting or joining.
Since each trace in the super Yang-Mills theory corresponds to a closed string state, this translates into the fact that,
in the planar limit in the super Yang-Mills theory, different multi-trace structures do not mix.
For the open string sector, when the string coupling goes to zero there is again no splitting and joining so that the open
string Chan-Paton factors are frozen.
The translation of a giant graviton system into an operator in the field theory is straight
forward\cite{Balasubramanian:2002sa,Balasubramanian:2004nb,Berenstein:2005fa,Berenstein:2006qk,dssi,dssii,bds,Koch:2011hb,deMelloKoch:2012ck}: 
in the operator $O^{(n,m)}_{r, j,j^3}$ each row of $r$ corresponds to a giant graviton and each impurity
$Y$ corresponds to an open string (this last interpretation is proved in \cite{Koch:2011hb,deMelloKoch:2012ck}).
$j^3$ tells us the number of open string end points attached to each giant.
Since the Chan-Paton factors are frozen, $j^3$ is not changed by the action of the dilatation operator and
\bea
    \beta(a,b,\pm 1)=0\, .
\eea
This now leaves 4 unknown terms to be determined.

Another consequence of working at large $N$ in the displaced corners approximation, is
\bea
  \beta^{(n,m)}_{r_1+\alpha ,j,j^3}(a,b,c)=  \beta^{(n,m)}_{r_1,j,j^3}(a,b,c)
\eea
with $\alpha$ any number of order 1. 
This follows because $r_1$ is order $N$ and the matrix elements of the dilatation operator depend smoothly on the parameters $r_1,j,j^3$, 
so we can replace $r_1+\alpha$ by $r_1$ making negligible error in the large $N$ limit. 
There is one point that deserves attention.
In general our results depend on $r_1$, $r_2$ and on $r_1-r_2$.
Even if $r_1=O(N)$ and $r_2=O(N)$, if $r_1-r_2=O(1)$, replacing $r_1+\alpha\to r_1$ can result in errors that do not vanish as $N\to\infty$.
In the displaced corners approximation all $r$ row lengths are well separated and this does not happen.
It then follows that the $r_i$ are conserved and that the coefficients of $\sqrt{r_1}$ and $\sqrt{r_2}$ in (\ref{rightalgebra}) must 
separately vanish. 
This has a very natural physical interpretation: the $r_i$ set the momenta of the giant gravitons and the back reaction 
on each giant graviton is negligible.

Although we are mainly interested in the dependence of the dilatation operator on $j,j^3$, we do know that
\bea
   \beta^{(n,m)}_{r_1 ,j,j^3}(\pm 1,d,e)&=&\sqrt{(N+r_1)(N+r_2)}f(j,j^3,d,e)\nn\\
                                        &=&\sqrt{(N+r_1)(N+n-r_1)}f(j,j^3,d,e)\nn\\
   \beta^{(n,m)}_{r_1 ,j,j^3}(0,d,e)&=&(2N+r_1+r_2)g(j,j^3,d,e)\nn\\
                                    &=&(2N+n)g(j,j^3,d,e)\,.\label{detailedansatz}
\eea
These formulas deserve some discussion.
The dependence of matrix elements on factors\footnote{Recall that a box in row $i$ and column $j$ 
          has a factor $N-i+j$.} of boxes in the Young diagram labels has two sources:

\begin{itemize}

\item[1.] There is an overall normalization $\sqrt{f_T\over f_R}$. The factors of any boxes that 
          are common to $R$ and $T$ will cancel so that we are left with
          \bea
            F_1=\sqrt{\prod_{i\in \textrm{boxes in $T$ that are not in $R$}}~~c_i\over\prod_{j\in \textrm{boxes in $R$ that are not in $T$}}~~c_j}
          \eea

\item[2.] When evaluating the dilatation operator, we need to sum over $S_{n+m}$. 
          As discussed above, derivatives with respect to $Y$ and $Z$ produce Kronecker delta functions that restrict the sum to the subgroup $S_{n+m-1}$. 
          The original trace over $R\vdash m+n$ then becomes a trace over an irreducible representation of the subgroup $R'\vdash m+n- 1$.
          The sum then produces the factor of the box that must be removed from $R$ to obtain $R'$.
          The trace splits into a trace over $r'\vdash n-1$ which sets $r'=t'$ and a trace over $s$ which depends only on $j,j^3$.
          This dependence is summarized in the functions $f(j,j^3,d,e)$ and $g(j,j^3,d,e)$ above and it is these functions that we want
          to constrain using the su(2) invariance.

\end{itemize}

For the first term in (\ref{detailedansatz}) we have
\bea
  F_1=\sqrt{N+r_1\over N+r_2}~~~F_2=N+r_2~~~~~\textrm{or}~~~~~ F_1=\sqrt{N+r_2\over N+r_1}~~~F_2=N+r_1
\eea
so that
\bea
   F_1\cdot F_2 =\sqrt{(N+r_1)(N+r_2)}
\eea
For the second term in (\ref{detailedansatz}) we have two contributions which both have $F_1=1$ and
\bea
    F_2=N+r_1~~~~~\textrm{or}~~~~~F_2=N+r_2
\eea
Thus, the total coefficient of this term is
\bea
   N+r_1+N+r_2=2N+r_1+r_2=2N+n
\eea

Since we are computing a commutator, the answer for $D_2$ will not be unique. 
Indeed, replacing $D_2\to D_2+\alpha {\bf 1}$ with $\alpha$ a constant, will not change the value of the commutator.
To fix the value of $\alpha$ note that there are BPS operators belonging to the $su(2)$ sector. 
These operators are annihilated by $D_2$, so that the smallest eigenvalue of $D_2$ is zero.
This fixes $\alpha$.

Now, use
\bea
  J_+ O^{(n,m)}_{r_1, j,j^3}=\sum_{a=-1}^0\sum_{b=-{1\over 2}}^{1\over 2}\alpha^{(n,m)}_{r_1 ,j,j^3}(a,b)
                             O^{(n-1,m+1)}_{r_1+a, j+b,j^3-{1\over 2}-a}
\eea
where (using the results of the last section)
\bea
   \alpha^{(n,m)}_{r_1,j,j^3}(-1,{1\over 2})=\sqrt{r_1}\sqrt{m+2j+4\over 2}{j+j^3+1\over\sqrt{(2j+2)(2j+1)}}
\eea
\bea
   \alpha^{(n,m)}_{r_1,j,j^3}(-1,-{1\over 2})=\sqrt{r_1}\sqrt{m-2j+2\over 2}{j-j^3\over\sqrt{2j(2j+1)}}
\eea
\bea
   \alpha^{(n,m)}_{r_1,j,j^3}(0,{1\over 2})=\sqrt{r_2}\sqrt{m+2j+4\over 2}{j-j^3+1\over\sqrt{(2j+2)(2j+1)}}
\eea
\bea
   \alpha^{(n,m)}_{r_1,j,j^3}(0,-{1\over 2})=\sqrt{r_2}\sqrt{m-2j+2\over 2}{j+j^3\over\sqrt{2j(2j+1)}}
\eea
and use
\bea
   D_2O^{(n,m)}_{r_1,j,j^3}=\sum_{a=1}^{-1}\sum_{b=1}^{-1}\sum_{c=1}^{-1}\beta^{(n,m)}_{r_1,j,j^3}(a,b,c)O^{(n,m)}_{r_1+a,j+b,j^3+c}
\eea
to evaluate
\bea
  \big[ J_+,D_2\big] O^{(n,m)}_{r_1,j,j^3}=0\, .
\eea
The result is
\bea
    \label{basicresult}
  &&\sum_{a=-1}^0\sum_{b=-{1\over 2}}^{1\over 2}\sum_{c=-1}^1\sum_{d=-1}^1\sum_{e=-1}^1
     \left(\beta^{(n,m)}_{r_1,j,j^3}(c,d,e)\alpha^{(n,m)}_{r_1+c,j+d,j^3+e}(a,b)\right.\\
  &&\left.-\alpha^{(n,m)}_{r_1,j,j^3}(a,b)\beta^{(n-1,m+1)}_{r_1+a,j+b,j^3-{1\over 2}-a}(c,d,e)\right)
   O^{(n-1,m+1)}_{r_1+a+c,j+d+b,j^3+e-{1\over 2}-a}=0\,.\nn
\eea
The operators $O^{(n,m)}_{r_1,j,j^3}$ are all linearly independent, so that the coefficient of each term must vanish separately.
Further, since $\alpha^{(n,m)}_{r_1,j,j^3}(-1,\cdot)\propto\sqrt{r_1}$ and $\alpha^{(n,m)}_{r_1,j,j^3}(0,\cdot)\propto\sqrt{r_2}$,
terms with different values of $a$ in $\alpha^{(n,m)}_{r_1,j,j^3}(a,\cdot)$ must separately vanish.

To illustrate some of the details, we will discuss some examples of equations that we obtain from (\ref{basicresult}).
In particular, we will explain how the $\beta_{r_1,j,j^3}(0,1,0)$ matrix element is determined.
Set $a=0$, $c=0$, $e=0$, $d+b=-{3\over 2}\Rightarrow (d,b)=(-1,-{1\over 2})$ to obtain
\bea
   \beta^{(n,m)}_{r_1,j,j^3}(0,-1,0)\alpha^{(n,m)}_{r_1,j-1,j^3}(0,-{1\over 2})
 - \alpha^{(n,m)}_{r_1,j,j^3}(0,-{1\over 2})\beta^{(n-1,m+1)}_{r_1,j-{1\over 2},j^3-{1\over 2}}(0,-1,0)=0\,,\nn
\eea
\bea
   \label{eqn12}
   &&\sqrt{m-2j+4\over 2}{j+j^3-1\over\sqrt{(2j-2)(2j-1)}}\beta^{(n,m)}_{r_1,j,j^3}(0,-1,0)\\
 &&- \sqrt{m-2j+2\over 2}{j+j^3\over\sqrt{2j(2j+1)}}\beta^{(n-1,m+1)}_{r_1,j-{1\over 2},j^3-{1\over 2}}(0,-1,0)=0\,.\nn
\eea
Next, set $a=-1$, $c=0$, $e=0$, $d+b=-{3\over 2}\Rightarrow (d,b)=(-1,-{1\over 2})$ to obtain
\bea
   \beta^{(n,m)}_{r_1,j,j^3}(0,-1,0)\alpha^{(n,m)}_{r_1,j-1,j^3}(-1,-{1\over 2})
 - \alpha^{(n,m)}_{r_1,j,j^3}(-1,-{1\over 2})\beta^{(n-1,m+1)}_{r_1-1,j-{1\over 2},j^3+{1\over 2}}(0,-1,0)=0\,,\nn
\eea
\bea
   \label{eqn13}
   &&\sqrt{m-2j+4\over 2}{j-j^3-1\over\sqrt{(2j-2)(2j-1)}}\beta^{(n,m)}_{r_1,j,j^3}(0,-1,0)\\
 &&- \sqrt{m-2j+2\over 2}{j-j^3\over\sqrt{(2j+1)2j}}\beta^{(n-1,m+1)}_{r_1-1,j-{1\over 2},j^3+{1\over 2}}(0,-1,0)=0\,.\nn
\eea
Combining (\ref{eqn12}) and (\ref{eqn13}) we find
\bea
  \label{goodone}
  \beta^{(n,m)}_{r_1,j,j^3}(0,-1,0)={j+j^3\over j+j^3-1}{j-j^3\over j-j^3+1}\beta^{(n,m)}_{r_1,j,j^3-1}(0,-1,0)
\eea
which implies that
\bea
   \beta^{(n,m)}_{r_1,j,j^3}(0,-1,0)\propto (j+j^3)(j-j^3)\, .
\eea
Daggering equation (\ref{goodone}) we find
\bea
  \beta^{(n,m)}_{r_1,j,j^3}(0,1,0)=
  {j+j^3+1\over j+j^3}{j-j^3+1\over j-j^3+2}\beta^{(n,m)}_{r_1,j,j^3-1}(0,1,0)
\eea
which implies that
\bea
   \beta^{(n,m)}_{r_1,j,j^3}(0,1,0)\propto (j+j^3+1)(j-j^3+1)\,.
\eea

Now, set $a=0$, $b={1\over 2}$, $c=0$, $d=1$ and $e=0$ to obtain
\bea
   \beta^{(n,m)}_{r_1,j,j^3}(0,1,0)\alpha^{(n,m)}_{r_1,j+1,j^3}(0,{1\over 2})
 - \alpha^{(n,m)}_{r_1,j,j^3}(0,{1\over 2})\beta^{(n-1,m+1)}_{r_1,j+{1\over 2},j^3-{1\over 2}}(0,1,0)=0\,,\nn
\eea
\bea
   \label{eqn14}
   &&\sqrt{m+2j+6\over 2}{j-j^3+2\over\sqrt{(2j+4)(2j+3)}}\beta^{(n,m)}_{r_1,j,j^3}(0,1,0)\\
 &&- \sqrt{m+2j+4\over 2}{j-j^3+1\over\sqrt{(2j+1)(2j+2)}}\beta^{(n-1,m+1)}_{r_1,j+{1\over 2},j^3-{1\over 2}}(0,1,0)=0\,.\nn
\eea
Daggering this we find
\bea
   \beta^{(n-1,m+1)}_{r_1,j-{1\over 2},j^3-{1\over 2}}(0,-1,0)=\sqrt{m+2j+2\over m+2j}{j-j^3\over j-j^3-1}
                       \sqrt{(2j-3)(2j-2)\over (2j-1)2j}\beta^{(n,m)}_{r_1,j-1,j^3}(0,-1,0)\,.\nn
\eea
Combining this with (\ref{eqn12}) we find
\bea
&&\beta^{(n,m)}_{r_1,j,j^3}(0,-1,0)=\sqrt{(m-2j+2)(m+2j+2)\over (m-2j+4)(m+2j)}{2j-2\over 2j}\sqrt{(2j-1)(2j-3)\over (2j+1)(2j-1)}
      \nn\\
&&\times{j+j^3\over j+j^3-1}{j-j^3\over j-j^3-1}\beta^{(n,m)}_{r_1,j-1,j^3}(0,-1,0)\nn
\eea
which implies that
\bea
   \beta^{(n,m)}_{r_1,j,j^3}(0,-1,0)\propto \sqrt{(m + 2j + 2)(m - 2j +2)\over (2j + 1)(2j-1)}
{(j+j^3 )(j-j^3 ) \over 2 j}
\label{dzmz}
\eea
which is indeed the correct result. Daggering, we find
\bea
   \beta^{(n,m)}_{r_1,j,j^3}(0,1,0)\propto\sqrt{(m + 2j + 4)(m - 2j)\over (2j + 1)(2j+3)} {(j+j^3+1)(j-j^3 + 1) \over 2(j + 1)}
\label{dzpz}
\eea
which is also correct.
Solving the complete set of recursion relations we find
\bea
&&D_2O^{(n,m)}(r_1,j,j^3)=\nn\\
&&\sqrt{(m-2j+2)(m+2j+2)\over (2j+1)(2j-1)}{(j+j^3)(j-j^3)\over 2j}
\left[c_{010}(2N+r_1+r_2)O^{(n,m)}(r_1,j-1,j^3)\right.\nn\\
&&\left. +c_{110}\sqrt{(N+r_1)(N+r_2)}(O^{(n,m)}(r_1-1,j-1,j^3)+O^{(n,m)}(r_1+1,j-1,j^3))\right]\nn\\
&&+\sqrt{(m+2j+4)(m-2j)\over (2j+3)(2j+1)}{(j+j^3+1)(j-j^3+1)\over 2j+2}
\left[c_{010}(2N+r_1+r_2)O^{(n,m)}(r_1,j+1,j^3)\right.\nn\\
&&\left. +c_{110}\sqrt{(N+r_1)(N+r_2)}(O^{(n,m)}(r_1-1,j+1,j^3)+O^{(n,m)}(r_1+1,j+1,j^3))\right]\nn\\
&&+\left(-{1\over 2}\left( m-{(m+2)(j^3)^2\over j(j+1)}\right)\right)
\left[c_{010}(2N+r_1+r_2)O^{(n,m)}(r_1,j,j^3)\right.\nn\\
&&\left. +c_{110}\sqrt{(N+r_1)(N+r_2)}(O^{(n,m)}(r_1-1,j,j^3)+O^{(n,m)}(r_1+1,j,j^3))\right]\nn
\eea
where $c_{010}$ and $c_{110}$ are arbitrary constants, independent of $j$, $j^3$ and $r_1$.
Thus, we have determined the $j$, $j^3$ dependence of the matrix elements of the one loop dilatation operator.
Achieving this at higher loops is one of the main goals of this article.
To completely determine the spectrum of anomalous dimensions, we need to determine the constants $c_{010}$ and $c_{110}$
in the above expression.
These constants are tightly constrained as we now explain.
In the large $N$ regime, we can take a continuum limit of the action of the dilatation operator.
Towards this end, introduce the continuous variable $\rho={r_1-r_2 \over 2\sqrt{N+r_2}}$ and replace $O^{(r,m)}(r_1,j,j^3)$ with $O^{(r,m)}(\rho,j,j^3)$.
$r_1$ is the longer (top) row and $r_2$ is the shorter bottom row. 
When $\rho$ is order 1 the dilatation operator becomes an $N$ independent differential operator\cite{gs}.
Expanding we have  
$$
\sqrt{(N+r_1)(N+r_2)}=(N+r_2)\left(
1+{1\over 2}{r_1-r_2\over N+r_2}-{1\over 8}{(r_1-r_2)^2\over (N+r_2)^2}+....\right)
$$
The first term above is $O(N)$, the second $O(\sqrt{N})$ and the third $O(1)$.
$$
O^{(n,m)}\left(\rho-{1\over\sqrt{N+r_2}},j,j^3\right)=O^{(n,m)}(\rho,j,j^3)
-{1\over\sqrt{N+r_2}}{\partial O^{(n,m)}\over\partial \rho}\Big|_{\rho,j,j^3}
+{1\over N+r_2}{\partial^2 O^{(n,m)}\over\partial \rho^2}\Big|_{\rho,j,j^3} +...
$$
These expansions are only valid if $r_1-r_2\ll N+r_2$, which is certainly not always the case. 
However, we will learn something about the relation between the coeficients $c_{110}$ and $c_{010}$ by studying this situation. 
Using these expansions we have
\bea
&&c_{010}(2N+r_1+r_2)O^{(n,m)}(r_1,j,j^3)\nn\cr
&&\qquad+c_{110}\sqrt{(N+r_1)(N+r_2)}(O^{(n,m)}(r_1-1,j,j^3)+O^{(n,m)}(r_1+1,j,j^3))\nn\\
&&=\big[c_{110}+2c_{010}\big](N+r_2)O^{(n,m)}(r_1,j,j^3)+{1\over 2}\big[c_{110}+2c_{010}\big]\sqrt{N+r_2}O^{(n,m)}(r_1,j,j^3)+O(1)\nn
\eea
Again, the lowest eigenvalue of this operator is zero, reflecting a BPS operator.
To achieve this, the $O(N)$ and $O(\sqrt{N})$ pieces of this expansion must cancel which determines $c_{110}+2c_{010}=0$.
Thus, up to an overall normalization which our argument can't determine, we have reproduced (\ref{basicdil}).

\section{Continuum Limit}

We have demonstrated that the requirement that the one loop dilatation operator closes the correct Lie algebra when commuted with an $su(2)$ subgroup 
of the ${\cal R}$-symmetry group determines a set of recursion relations.
Solving these recursion relations we have recovered the formula for the one loop dilatation operator derived in \cite{Carlson:2011hy,Koch:2011hb} by
detailed computation.
We are interested in carrying this analysis out at higher loops.
The resulting recursion relations become very clumsy to solve.
To overcome this difficulty, we will now pursue a continuum approach to the problem, replacing the discrete variables $j,j^3$ by continuous
variables $x_j,x_{j^3}$.
The advantage of considering a continuum limit is that our recursion relations will be replaced by partial differential equations and we are
able to explicitely determine the general solution of these partial differential equations.
In this section we will motivate the continuum limit we study by considering the dilatation operator eigenproblem at one loop. 

The structure of the action of the one loop dilatation operator problem given in (\ref{basicdil}) exhibits an interesting factorization.
There is an action of $\Delta$ which acts only on the $r$ label times an action that is only on the $j,j^3$ labels.
The continuum limit we consider here is concerned with the action on the $j,j^3$ labels.
Recall that we take $m$ to be $O(\sqrt{N})$.
The discrete eigenproblem that we consider is\cite{Carlson:2011hy,Koch:2011hb}
\bea
  -\lambda \psi (j,j^3)=
  \sqrt{(m + 2j + 4)(m - 2j)\over (2j + 1)(2j+3)} {(j+j^3 + 1)(j-j^3 + 1) \over 2(j + 1)} \psi (j+1,j^3)\cr
  \sqrt{(m + 2j + 2)(m - 2j +2)\over (2j + 1)(2j-1)} {(j+j^3 )(j-j^3 ) \over 2 j}\psi (j-1,j^3)
  -{1\over 2}\left( m-{(m+2)(j^3)^2\over j(j+1)}\right)\psi (j,j^3)\, .\cr
\eea
The variables that become continuous as we take $N\to\infty$ are
\bea
   x_j={j\over\sqrt{m}},\qquad x_{j^3}={j^3\over\sqrt{m}}
   \label{twocontvars}
\eea
Replace $\psi (j,j^3)$ by $\psi (x_j,x_{j^3})$ and use the expansions
\bea
   -{1\over 2}\left( m-{(m+2)(j^3)^2\over j(j+1)}\right)= -{m\over 2}+{m\over 2}{x^2_{j^3}\over x^2_j}
                   -{\sqrt{m}\over 2}{x_{j^3}^2\over x_j^3}+{x_{j^3}^2\over 2x_j^4}+{x_{j^3}^2\over x_j^2}
\eea
\bea
\sqrt{(m+2j+4)(m-2j)\over (2j+1)(2j+3)}{(j+j^3+1)(j-j^3+1)\over 2(j+1)}={m\over 4}+{1\over 2}-{x_j^2\over 2}+{1\over 32x_j^2}\cr
-{m\over 4}{x_{j^3}^2\over x_j^2}-{1\over 2}{x_{j^3}^2\over x_j^2}+{x_{j^3}^2\over 2}-{25x_{j^3}^2\over 32x_j^4}
+{\sqrt{m}x_{j^3}^2\over 2 x_j^3}
\eea
\bea
\sqrt{(m + 2j+2)(m - 2j+2)\over (2j + 1)(2j-1)} {(j+j^3 )(j-j^3 ) \over 2 j}
={m\over 4}+{1\over 2}-{x_j^2\over 2}+{1\over 32x_j^2}\cr
-{m\over 4}{x_{j^3}^2\over x_j^2}-{1\over 2}{x_{j^3}^2\over x_j^2}+{x_{j^3}^2\over 2}-{x_{j^3}^2\over 32x_j^4}\,.
\eea
It is now a simple matter to find the following eigenproblem in the continuum
\bea
{1\over 4}\left(1-{x_{j^3}^2\over x_j^2}\right){d^2 \psi \over dx_j^2}
+{x_{j^3}^2\over 2x_j^3}{d \psi \over dx_j}
+\left[
-{5x_{j^3}^2\over 16x_j^4}+1-x_j^2+{1\over 16x_j^2}+x_{j^3}^2
\right]\psi =-\lambda \psi \,.
\eea
In obtaining this result the form for our continuum limit, as spelled out in (\ref{twocontvars}) is crucial.
Indeed, if one sets $x_j=j/m^\alpha$ the ``kinetic'' and ``harmonic potential'' terms on the LHS are only the same size if $\alpha={1\over 2}$. 
Now, set $\psi =\sqrt{x_j}g$ to obtain
\bea
{1\over 4}\left(1-{x_{j^3}^2\over x_j^2}\right)
{d^2 g\over dx_j^2}+{1\over 4x_j}\left(1+{x_{j^3}^2\over 2x_j^3}\right){dg\over dx_j}+
\left[
1-x_j^2+x_{j^3}^2
\right]g=-\lambda g\,.
\eea
Finally, in terms of the new variable $u$ defined by $u^2=x_j^2-x_{j^3}^2$ we find
\bea
  {1\over 4}{d^2 g\over du^2}+{1\over 4u}{dg\over du}+(1-u^2)g=-\lambda g\, .
  \label{conteigprob}
\eea
If we set $r=2u$ we find the eigenproblem of the 2-dimensional oscillator with zero angular momentum.
The energy spacing is 2 (recall $j\ge 0$ to see this).
This is exactly the spectrum obtained by solving the discrete problem\cite{Carlson:2011hy,Koch:2011hb}.
It is also easy to check that the eigenvectors of the discrete problem are in perfect agreement with the eigenfunctions of (\ref{conteigprob}).
Thus, the continuum problem contains the same information as the discrete problem.

To get the correct spectrum we must obtain the $O(m)$, $O(\sqrt{m})$ and $O(1)$ pieces of the matrix elements of the dilatation operator.
Writing things schematically, we should expand our dilatation operator matrix elements as 
\bea
   \beta = mf^{(0)}+\sqrt{m}f^{(1)}+f^{(2)}+{f^{(3)}\over\sqrt{m}}+O({1\over m})
\eea
and we should expand 
\bea
   \alpha = \sqrt{m}\alpha^{(0)} +\alpha^{(1)} +{1\over\sqrt{m}}\alpha^{(2)}+{1\over m}\alpha^{(3)}+O({1\over m^{3\over 2}})
\eea
After expansion (\ref{rightalgebra}) gives 3 sets of non-trivial equations, and these three equations are the
complete content of the recursion relations.
They are obtained by plugging the above expansions into (\ref{rightalgebra}) and setting the coefficients of $m$, $\sqrt{m}$ and 1 to zero.
The terms with coefficient $m^{3\over 2}$ trivially vanish. 
The terms with negative powers of $m$ also do not give new equations: they vanish automatically because we are working in 
the $m=\sqrt{N}\to\infty$ limit. 

At one loop, solving the partial differential equations that arise from (\ref{rightalgebra}) must reproduce the following expansions
\bea
\beta^{(n,m)}_{r_1,j,j^3}(c,0,0)=-{m\over 2}+{m\over 2}{x^2_{j^3}\over x^2_j}
                   -{\sqrt{m}\over 2}{x_{j^3}^2\over x_j^3}+{x_{j^3}^2\over 2x_j^4}+{x_{j^3}^2\over x_j^2}
\eea
\bea
\beta^{(n,m)}_{r_1,j,j^3}(c,1,0)={m\over 4}+{1\over 2}-{x_j^2\over 2}+{1\over 32x_j^2}
-{m\over 4}{x_{j^3}^2\over x_j^2}-{1\over 2}{x_{j^3}^2\over x_j^2}+{x_{j^3}^2\over 2}-{25x_{j^3}^2\over 32x_j^4}
+{\sqrt{m}x_{j^3}^2\over 2 x_j^3}
\eea
\bea
\beta^{(n,m)}_{r_1,j,j^3}(c,-1,0)={m\over 4}+{1\over 2}-{x_j^2\over 2}+{1\over 32x_j^2}
-{m\over 4}{x_{j^3}^2\over x_j^2}-{1\over 2}{x_{j^3}^2\over x_j^2}+{x_{j^3}^2\over 2}-{x_{j^3}^2\over 32x_j^4}
\eea
Given these continuum results, we can immediately claim that we have reproduced (\ref{basicdil}).
Indeed, the ambiguity in reconstructing the exact functions $\beta^{(n,m)}_{r_1,j,j^3}(c,d,0)$ of the discrete variables $j,j^3$
from the continuum expressions above is order ${1\over m}$ and we are working in the $m=\sqrt{N}\to\infty$ limit.

Finally, it is important to note that the solutions to our continuum differential equations are not unique.
Indeed, we are finding a dilatation operator $D$ that obeys
\bea
   \left[ J_{\pm},D\right]=0=\left[J_3,D\right]\, .
\eea
Given a first solution, another solution is easily constructed by rescaling and shifting
\bea
   D\to \kappa_1 D + 2k_0{\bf 1}
\eea 
where ${\bf 1}$ is the identity. 
Thus, there will always be two arbitrary constants in our solutions.
This has important implications for us, particularly when it comes to finding the most general solution to the partial differential equations
we will derive.
For example, by choosing $\kappa_1={1\over \sqrt{m}}\gamma$ we see that we shift
\bea
  \beta = mf^{(0)}+\sqrt{m}f^{(1)}+f^{(2)}+{f^{(3)}\over\sqrt{m}}+O({1\over m})
\longrightarrow\nn\cr
  \beta' = mf^{(0)}+\sqrt{m}(f^{(1)}+\gamma f^{(0)})+f^{(2)}+\gamma f^{(1)}+{f^{(3)}+\gamma f^{(2)}\over\sqrt{m}}+O({1\over m})\nn
\eea
In what follows, we will construct the solution that has $\gamma =0$ and say that ``we have the most general solution up to symmetry''.
Note that by choosing $\kappa_1={1\over m}\gamma$ we would have
\bea
  \beta' = mf^{(0)}+\sqrt{m}f^{(1)}+f^{(2)}+\gamma f^{(0)}+{f^{(3)}+\gamma f^{(1)}\over\sqrt{m}}+O({1\over m})\, .\nn
\eea
We will thus also not include terms $\propto f^{(0)}$ when solving the partial differential equations that determine $f^{(2)}$.
This completes out discussion of the continuum limit.

\section{Differential Equations and Higher Loop Anomalous Dimensions}\label{higherloopanalysis}

The main goal of this section is to study the constraints implied by (\ref{rightalgebra}) on the $p$-loop dilatation operator.
As we discussed above, the $p$-loop dilatation operator allows a total of $p$ boxes on the Young diagram labels of the restricted
Schur polynomial to move. 
In this case, the requirement that $J_+$ commutes with $D$ implies that
\bea
   \sum_{b=-{1\over 2}}^{1\over 2}\sum_{d=-p}^p\left[
   \beta^{(n,m)}_{r_1,j,j^3}(c,d,0)\alpha^{(n,m)}_{r_1+c,j+d,j^3}(a,b)\right.\cr
   \left. -\alpha^{(n,m)}_{r_1,j,j^3}(a,b)\beta^{(n-1,m+1)}_{r_1+a,j+b,j^3-{1\over 2}-a}(c,d,0)\right]
   O^{(n-1,m+1)}_{r_1+a+c,j+d+b,j^3-{1\over 2}-a}=0
   \label{basicreln}
\eea
which can be rewritten as
\bea
&&\,\,\, \beta^{(n,m)}_{r_1,j,j^3}(c,d,0)\alpha^{(n,m)}_{r_1+c,j+d,j^3}(a,{1\over 2})
  +\beta^{(n,m)}_{r_1,j,j^3}(c,d+1,0)\alpha^{(n,m)}_{r_1+c,j+d+1,j^3}(a,-{1\over 2})\cr
&&-\alpha^{(n,m)}_{r_1,j,j^3}(a,{1\over 2})\beta^{(n-1,m+1)}_{r_1+a,j+{1\over 2},j^3-{1\over 2}-a}(c,d,0)
  -\alpha^{(n,m)}_{r_1,j,j^3}(a,-{1\over 2})\beta^{(n-1,m+1)}_{r_1+a,j-{1\over 2},j^3-{1\over 2}-a}(c,d+1,0)=0\, .\cr
\label{starteqn}\eea
Recall that $\alpha (-1,\cdot)\propto\sqrt{r_1}$ and $\alpha (0,\cdot)\propto\sqrt{r_2}$ so that we get independent equations from 
(\ref{basicreln}) for each value of $a=\{-1,0\}$, $c=\{-p,-p+1,\cdots,p-1,p\}$, and $d+b$ where $b=\pm{1\over 2}$ and $d=\{-p,-p+1,\cdots,p-1,p\}$.
We will freely make use of the result of the Appendix in this section.

To begin we will consider $a=0$ in (\ref{starteqn}). 
A few words on how we perform the expansion of the $\alpha_{r_1,j,j^3}(a,\pm {1\over 2})$ is in order.
After rewriting $j,j^3$ in terms of $x_j,x_{j^3}$
\bea
&&\alpha_{r_1,j,j^3}(0,{1\over 2})=\sqrt{r_2}\sqrt{m+2j+4\over 2}{j-j^3+1\over\sqrt{2j+2}\sqrt{2j+1}}\cr
&&=\sqrt{r_2}\sqrt{m\over 2}\sqrt{1+2{x_j\over \sqrt{m}}+{4\over m}}
{x_j-x_{j^3}+{1\over \sqrt{m}}\over\sqrt{2x_j+{2\over\sqrt{m}}}\sqrt{2x_j+{1\over\sqrt{m}}}}
\eea
we perform an expansion treating ${1\over\sqrt{m}}$ and ${1\over m}$ as small numbers.
Using these expansions, after equating the coefficients of $m^{3\over 2}$ to zero, in
\bea
 \beta^{(n,m)}_{r_1,j,j^3}(c,d,0)                          \alpha^{(n,m)}_{r_1+c,j+d,j^3}(0,{1\over 2})
-\beta^{(n-1,m+1)}_{r_1,j+{1\over 2},j^3-{1\over 2}}(c,d,0)\alpha^{(n,m)}_{r_1,j,j}(0,{1\over 2})\cr
+\beta^{(n,m)}_{r_1,j,j^3}(c,d+1,0)\alpha^{(n,m)}_{r_1+c,j+d+1,j^3}(0,-{1\over 2})
-\beta^{(n-1,m+1)}_{r_1,j-{1\over 2},j^3-{1\over 2}}(c,d+1,0)\alpha^{(n,m)}_{r_1,j,j^3}(0,-{1\over 2})
=0
\label{forfirst}
\eea
we find
\bea
&& \frac{(x_j-x_{j^3})}{2 \sqrt{2} x_j}f^{(0)}_{c,d}(x_j,x_{j^3})
  +\frac{(x_j+x_{j^3})}{2 \sqrt{2} x_j}f^{(0)}_{c,d+1}(x_j,x_{j^3})\cr
&&-\frac{(x_j-x_{j^3})}{2\sqrt{2} x_j}f^{(0)}_{c,d}(x_j,x_{j^3})
  -\frac{(x_j+x_{j^3})}{2 \sqrt{2} x_j}f^{(0)}_{c,d+1}(x_j,x_{j^3})=0
\eea
which is trivially obeyed. By equating the $O(m)$ term to zero we have
\bea
       2 x_{j^3} (d f^{(0)}_{c,d}-(d+1) f^{(0)}_{c,d+1})
       +x_j \left(x_j \left({\partial f^{(0)}_{c,d}\over \partial x_{j^3}}
                 +{\partial f^{(0)}_{c,d+1}\over \partial x_{j^3}}
                 -{\partial f^{(0)}_{c,d}\over \partial x_j}
                 +{\partial f^{(0)}_{c,d+1}\over \partial x_j}\right)\right.\cr
                 \left. +x_{j^3}
                \left(-{\partial f^{(0)}_{c,d}\over \partial x_{j^3}}
                 +{\partial f^{(0)}_{c,d+1}\over \partial x_{j^3}}
                 +{\partial f^{(0)}_{c,d}\over \partial x_j}
                 +{\partial f^{(0)}_{c,d+1}\over \partial x_j}\right)\right)=0
\label{firsttouse}
\eea
Equating the $O(\sqrt{m})$ term to zero gives
\bea
x_j \left(2 d \left(4 x_j^2-1\right) (f^{(0)}_{c,d}-f^{(0)}_{c,d+1}) 
-x_j \left[ x_j 
   \left({\partial^2 f^{(0)}_{c,d}\over\partial x_{j^3}^2}
   +{\partial^2 f^{(0)}_{c,d+1}\over\partial x_{j^3}^2}
-4 {\partial f^{(1)}_{c,d}\over\partial x_{j^3}}-4
   {\partial f^{(1)}_{c,d+1}\over\partial x_{j^3}}\right.\right.\right.\cr
\left.\left.-2 {\partial^2 f^{(0)}_{c,d}\over\partial x_j\partial x_{j^3}}
+2 {\partial^2 f^{(0)}_{c,d+1}\over\partial x_j\partial x_{j^3}}\right)
-{\partial f^{(0)}_{c,d}\over\partial x_{j^3}}
+{\partial f^{(0)}_{c,d+1}\over\partial x_{j^3}}
+{\partial f^{(0)}_{c,d}\over\partial x_j}
+{\partial f^{(0)}_{c,d+1}\over\partial x_j}\right]+x_j^2 \left(-\left[4 x_j
   \left(-{\partial f^{(0)}_{c,d}\over\partial x_{j^3}}\right.\right.\right.\cr
\left.+{\partial f^{(0)}_{c,d+1}\over\partial x_{j^3}}
+{\partial f^{(0)}_{c,d}\over\partial x_j}
+{\partial f^{(0)}_{c,d+1}\over\partial x_j}\right)+8 f^{(0)}_{c,d}+16 f^{(0)}_{c,d+1}
+4 {\partial f^{(1)}_{c,d}\over\partial x_j}-4
   {\partial f^{(1)}_{c,d+1}\over\partial x_j}
+{\partial^2 f^{(0)}_{c,d}\over\partial x_j^2}\cr
\left.\left.\left.
+{\partial^2 f^{(0)}_{c,d+1}\over\partial x_j^2}\right]\right)
+2 f^{(0)}_{c,d+1}\right)-x_{j^3} \left(x_j \left[-8 d
 f^{(1)}_{c,d}
+8 d f^{(1)}_{c,d+1}-3
   {\partial f^{(0)}_{c,d}\over\partial x_{j^3}}
+{\partial f^{(0)}_{c,d+1}\over\partial x_{j^3}}
+8 f^{(1)}_{c,d+1}
\right.\right.\cr
\left.
+3 {\partial f^{(0)}_{c,d}\over\partial x_j}
+{\partial f^{(0)}_{c,d+1}\over\partial x_j}\right] +4(2d+3)\left[d f^{(0)}_{c,d}-(d+1) f^{(0)}_{c,d+1}\right] +x_j^2
   \left[-{\partial^2 f^{(0)}_{c,d}\over\partial x_{j^3}^2}
+{\partial^2 f^{(0)}_{c,d+1}\over\partial x_{j^3}^2}
+4 {\partial f^{(1)}_{c,d}\over\partial x_{j^3}}\right.\cr
-4 {\partial f^{(1)}_{c,d+1}\over\partial x_{j^3}}
+2 {\partial^2 f^{(0)}_{c,d}\over\partial x_j\partial x_{j^3}}
+2 {\partial^2 f^{(0)}_{c,d+1}\over\partial x_j\partial x_{j^3}}
-8 f^{(0)}_{c,d}+8 f^{(0)}_{c,d+1}
-4 {\partial f^{(1)}_{c,d}\over\partial x_j}
-4 {\partial f^{(1)}_{c,d+1}\over\partial x_j}\cr
\left. -{\partial^2 f^{(0)}_{c,d}\over\partial x_j^2}
+{\partial^2 f^{(0)}_{c,d+1}\over\partial x_j^2}\right]+4 x_j^3
\left.\left({\partial f^{(0)}_{c,d}\over\partial x_{j^3}}
+{\partial f^{(0)}_{c,d+1}\over\partial x_{j^3}}
-{\partial f^{(0)}_{c,d}\over\partial x_j}
+{\partial f^{(0)}_{c,d+1}\over\partial x_j}\right)\right)=0\cr
\label{FFeqn}
\eea
Finally, equating the $O(1)$ term to zero we find another equation that is rather long and hence we will not quote it here.
We will also study the equations obtained by plugging $a=-1$ into (\ref{starteqn}). 
Equating the term in
\bea
 \beta^{(n,m)}_{r_1,j,j^3}(c,d,0)                          \alpha^{(n,m)}_{r_1+c,j+d,j^3}(-1,{1\over 2})
-\beta^{(n-1,m+1)}_{r_1-1,j+{1\over 2},j^3+{1\over 2}}(c,d,0)\alpha^{(n,m)}_{r_1,j,j}(-1,{1\over 2})\cr
+\beta^{(n,m)}_{r_1,j,j^3}(c,d+1,0)\alpha^{(n,m)}_{r_1+c,j+d+1,j^3}(-1,-{1\over 2})
-\beta^{(n-1,m+1)}_{r_1-1,j-{1\over 2},j^3+{1\over 2}}(c,d+1,0)\alpha^{(n,m)}_{r_1,j,j^3}(-1,-{1\over 2})
=0\cr
\label{forsecond}
\eea
of order $m^{3/2}$ to zero, we find the equation 
\bea
 \frac{(x_j+x_{j^3})}{2 \sqrt{2} x_j}f^{(0)}_{c,d}(x_j,x_{j^3})
+\frac{(x_j-x_{j^3})}{2 \sqrt{2} x_j}f^{(0)}_{c,d+1}(x_j,x_{j^3})
\cr
-\frac{(x_j+x_{j^3})}{2\sqrt{2} x_j}f^{(0)}_{c,d}(x_j,x_{j^3})
-\frac{(x_j-x_{j^3})}{2 \sqrt{2} x_j}f^{(0)}_{c,d+1}(x_j,x_{j^3})=0
\eea
that is again trivially obeyed. The coefficient of the term of order $m$ is
\bea
  -x_j^2\left({\partial f^{(0)}_{c,d}\over \partial x_{j^3}}
+{\partial f^{(0)}_{c,d+1}\over \partial x_{j^3}}
+{\partial f^{(0)}_{c,d}\over \partial x_j}
-{\partial f^{(0)}_{c,d+1}\over \partial x_j}\right)\cr
 -x_{j^3} \left(2 d f^{(0)}_{c,d}-2 (d+1) f^{(0)}_{c,d+1}
+x_j
   \left(
    {\partial f^{(0)}_{c,d}\over \partial x_{j^3}}
   -{\partial f^{(0)}_{c,d+1}\over \partial x_{j^3}}
   +{\partial f^{(0)}_{c,d}\over \partial x_j}
   +{\partial f^{(0)}_{c,d+1}\over \partial x_j}\right)\right)=0
\label{secondtouse}
\eea
From the coefficient of the $O(\sqrt{m})$ term we find
\bea
x_j \left(-x_j^2 \left[-8 (d-1) f^{(0)}_{c,d}+8 (d+2) f^{(0)}_{c,d+1}
+{\partial^2 f^{(0)}_{c,d}\over\partial x_{j^3}^2}
+{\partial^2 f^{(0)}_{c,d+1}\over\partial x_{j^3}^2}
+4 {\partial f^{(1)}_{c,d}\over\partial x_{j^3}}
+4 {\partial f^{(1)}_{c,d+1}\over\partial x_{j^3}}\right.\right.\cr
\left.
+2 {\partial^2 f^{(0)}_{c,d}\over\partial x_j\partial x_{j^3}}
-2 {\partial^2 f^{(0)}_{c,d+1}\over\partial x_j\partial x_{j^3}}
+4 {\partial f^{(1)}_{c,d}\over\partial x_j}
-4 {\partial f^{(1)}_{c,d+1}\over\partial x_j}
+{\partial^2 f^{(0)}_{c,d}\over\partial x_j^2}
+{\partial^2 f^{(0)}_{c,d+1}\over\partial x_j^2}\right]\cr
+2 (d (f^{(0)}_{c,d+1}-f^{(0)}_{c,d})+f^{(0)}_{c,d+1})-4 x_j^3 \left[
 {\partial f^{(0)}_{c,d}\over\partial x_{j^3}}
-{\partial f^{(0)}_{c,d+1}\over\partial x_{j^3}}
+{\partial f^{(0)}_{c,d}\over\partial x_j}
+{\partial f^{(0)}_{c,d+1}\over\partial x_j}\right]
\cr
\left.
-x_j\left[
 {\partial f^{(0)}_{c,d}\over\partial x_{j^3}}
-{\partial f^{(0)}_{c,d+1}\over\partial x_{j^3}}
+{\partial f^{(0)}_{c,d}\over\partial x_j}
+{\partial f^{(0)}_{c,d+1}\over\partial x_j}\right]\right)
-x_{j^3} \left(-x_j \left[-8 d
 f^{(1)}_{c,d}+8 d
 f^{(1)}_{c,d+1}+3
   {\partial f^{(0)}_{c,d}\over\partial x_{j^3}}
\right.\right.\cr
\left.
-{\partial f^{(0)}_{c,d+1}\over\partial x_{j^3}}+8
 f^{(1)}_{c,d+1}
+3{\partial f^{(0)}_{c,d}\over\partial x_j}
+{\partial f^{(0)}_{c,d+1}\over\partial x_j}\right]
-4 (2 d+3) (d f^{(0)}_{c,d}-(d+1) f^{(0)}_{c,d+1})
\cr
+x_j^2
   \left({\partial^2 f^{(0)}_{c,d}\over\partial x_{j^3}^2}
-{\partial^2 f^{(0)}_{c,d+1}\over\partial x_{j^3}^2}
+4 {\partial f^{(1)}_{c,d}\over\partial x_{j^3}}
-4 {\partial f^{(1)}_{c,d+1}\over\partial x_{j^3}}
+2 {\partial^2 f^{(0)}_{c,d}\over\partial x_j\partial x_{j^3}}
+2 {\partial^2 f^{(0)}_{c,d+1}\over\partial x_j\partial x_{j^3}}
 +8 f^{(0)}_{c,d}-8 f^{(0)}_{c,d+1}\right.
\cr
\left.
+4 {\partial f^{(1)}_{c,d}\over\partial x_j}
+4 {\partial f^{(1)}_{c,d+1}\over\partial x_j}
+{\partial^2 f^{(0)}_{c,d}\over\partial x_j^2}
-{\partial^2 f^{(0)}_{c,d+1}\over\partial x_j^2}\right)+4 x_j^3\left(
 {\partial f^{(0)}_{c,d}\over\partial x_{j^3}}
+{\partial f^{(0)}_{c,d+1}\over\partial x_{j^3}}
\left. 
+{\partial f^{(0)}_{c,d}\over\partial x_j}
-{\partial f^{(0)}_{c,d+1}\over\partial x_j}\right)\right)=0\cr
\label{SSeqn}
\eea
Finally, the coefficient of the $O(1)$ term gives another long equation that we will again not quote.

Apart from the partial differential equations obtained above, we also need to require that the dilatation operator is hermittian.
Recall that
\bea
   (\beta^\dagger)^{(n,m)}_{r_1,j,j^3}(a,b,c)=\beta^{(n,m)}_{r_1+c,j+b,j^3+c}(-a,-b,-c)
\eea
Thus, we require
\bea
   \beta^{(n,m)}_{r_1,j,j^3}(c,q,0)=\beta^{(n,m)}_{r_1,j+q,j^3}(-c,-q,0)=\beta^{(n,m)}_{r_1,j+q,j^3}(c,-q,0)
\eea
which implies that
\bea
             &&m f^{(0)}_{c,a}(x_j,x_{j^3})
       +\sqrt{m} f^{(1)}_{c,a}(x_j,x_{j^3})
                +f^{(2)}_{c,a}(x_j,x_{j^3})
+{1\over\sqrt{m}}f^{(3)}_{(c,a)}(x_j,x_{j^3})\cr
    &&=        m f^{(0)}_{c,-a}(x_j+{a\over\sqrt{m}},x_{j^3})
       +\sqrt{m} f^{(1)}_{c,-a}(x_j+{a\over\sqrt{m}},x_{j^3})
               + f^{(2)}_{c,-a}(x_j+{a\over\sqrt{m}},x_{j^3})\cr
&&
+{1\over\sqrt{m}}f^{(3)}_{(c,-a)}(x_j+{a\over\sqrt{m}},x_{j^3})\, .
\label{hermitticityconstraint}
\eea

Our goal now is to solve the equations given above for the leading order of the functions introduced. 
There are two equations we will use: (\ref{firsttouse}) and (\ref{secondtouse}).
Introduce the functions
\bea
   F_+\equiv f^{(0)}_{c,d}+f^{(0)}_{c,d+1}\qquad   F_-\equiv f^{(0)}_{c,d}-f^{(0)}_{c,d+1}
\eea
In terms of these functions (\ref{firsttouse}) becomes
\bea
  2x_{j^3}\left[dF_- +{F_--F_+\over 2}\right]
+x_j^2\left[{\partial F_+\over\partial x_{j^3}}-{\partial F_-\over\partial x_j}\right]
+x_j x_{j^3}\left[{\partial F_+\over\partial x_j}-{\partial F_-\over\partial x_{j^3}}\right]=0
\label{impeqn}
\eea
and (\ref{secondtouse}) becomes
\bea
  2x_{j^3}\left[dF_- +{F_--F_+\over 2}\right]
+x_j^2\left[{\partial F_+\over\partial x_{j^3}}+{\partial F_-\over\partial x_j}\right]
+x_j x_{j^3}\left[{\partial F_+\over\partial x_j}+{\partial F_-\over\partial x_{j^3}}\right]=0\, .
\eea
Suming these two equations we learn that
\bea
  x_j{\partial F_-\over\partial x_j}+x_{j^3}{\partial F_-\over\partial x_{j^3}}=0
\eea
which implies that
\bea
   F_{-}=F_-(u)\qquad u={x_{j^3}\over x_j}\, .
\eea
Note that this holds for any $d$. 
If we set $d=p$, since $F_-=f^{(0)}_{c,p}$ we learn that $f^{(0)}_{c,p}=f^{(0)}_{c,p}(u)$.
If we set $d=p-1$, since $F_-=f^{(0)}_{c,p-1}-f^{(0)}_{c,p}$ depends only on $u$ and we already argued that $f^{(0)}_{c,p}$ depends
only on $u$, we learn that $f^{(0)}_{c,p-1}=f^{(0)}_{c,p-1}(u)$.
We can keep going in this way and consequently we have actually proved that
\bea
   f^{(0)}_{cd}=f^{(0)}_{cd}(u)
\eea
for {\it any} $d$.
This is a dramatic simplification - we had a collection of functions of two variables and now
we have a collection of functions that depend only on one variable.

Now, again set $d=p$. In this case $F_+=F_-=F(u)$. We find that (\ref{firsttouse}) becomes
\bea
  x_j{\partial F\over\partial x_{j^3}}+x_{j^3}{\partial F\over\partial x_j}=-2{x_{j^3}\over x_j}pF
\eea
which has the general solution
\bea
   F=f^{(0)}_{c,p}=\kappa_p (1-u^2)^p= \kappa_p \left( 1-{x_{j^3}^2\over x_j^2}\right)^p
   \label{ploopone}
\eea
where $\kappa_p$ is a constant.
This has reproduced the correct answer for one loop when $p=1$ and has determined the
leading order to an infinite number of higher loop dilatation operator coefficients.

Now, return to (\ref{impeqn}), and rewrite it using the new variable $y=1-u^2$ to obtain the simple form
\bea
y{df^{(0)}_{c,d}  \over dy}+
y{df^{(0)}_{c,d+1}\over dy}
=
df^{(0)}_{c,d} -(d+1)df^{(0)}_{c,d+1}\, .
\label{simplest}
\eea
If we now, set $d=p-1$ in (\ref{simplest}) we can solve to obtain
\bea
 f^{(0)}_{c,p-1} =-2p\kappa_p y^p + \kappa_{p-1}y^{p-1}\, .
\label{plooptwo}
\eea
Next, set $d=p-2$ in (\ref{simplest}) and again solve to obtain
\bea
 f^{(0)}_{c,p-2} =p(2p-1)\kappa_p y^p -2(p-1)\kappa_{p-1} y^{p-1} + \kappa_{p-2}y^{p-2}\, .
\eea
It is clear that we could continue with this process and determine all of the $ f^{(0)}_{c,d}$.
We have however determined all that we will need about the leading order.
We will now show that we can determine the one loop answer and then return to the general $p$-loop analysis.

\subsection{One Loop}

To determine the next to leading order, plug $d=1$ and the known leading order functions into (\ref{FFeqn}) to obtain
\bea
 {2\over x_{j^3}} \left(
   {\partial f^{(1)}_{c,1}\over \partial x_{j^3}} 
  -{\partial f^{(1)}_{c,1}\over \partial x_j}\right)
  x_j^4 (x_j-x_{j^3})+4 f^{(1)}_{c,1} x_j^3-\kappa_1 (2 x_j-x_{j^3})
   (x_j+x_{j^3}) =0\,,\cr
\label{firstd=1}
\eea
plug $d=0$ and the known leading order functions into (\ref{FFeqn}) to obtain
\bea
 x_j^3 x_{j^3} \left(x_j \left[
-{\partial f^{(1)}_{c,0}\over\partial x_{j^3}}
+{\partial f^{(1)}_{c,1}\over\partial x_{j^3}}
+{\partial f^{(1)}_{c,0}\over\partial x_j}
+{\partial f^{(1)}_{c,1}\over\partial x_j}\right]-2 f^{(1)}_{c,1}\right)\cr
 +x_j^5
\left({\partial f^{(1)}_{c,0}\over\partial x_{j^3}}
+{\partial f^{(1)}_{c,1}\over\partial x_{j^3}}
-{\partial f^{(1)}_{c,0}\over\partial x_j}
+{\partial f^{(1)}_{c,1}\over\partial x_j}\right)+\kappa_1 x_j x_{j^3}^2+\kappa_1 x_{j^3}^3=0\,,
\label{firstd=0}
\eea
and finally, plug $d=-2$ and the known leading order functions into (\ref{FFeqn}) to obtain
\bea
  x_j \left({\partial f^{(1)}_{c,-1}\over\partial x_{j^3}}
+{\partial f^{(1)}_{c,-1}\over\partial x_j}\right) (x_j+x_{j^3})+2 f^{(1)}_{c,-1} x_{j^3}=0\, .
\label{firstd=-1}
\eea
Next, plug $d=1$ and the known leading order functions into (\ref{SSeqn}) to obtain
\bea
 -{2\over x_{j^3}}\left({\partial f^{(1)}_{c,1}\over \partial x_{j^3}} +
          {\partial f^{(1)}_{c,1}\over \partial x_j}\right)
x_j^4 (x_j+x_{j^3})
-\left(4 f^{(1)}_{c,1} x_j^3+\kappa_1 \left(-2 x_j^2+x_j x_{j^3}+x_{j^3}^2\right)\right)=0\,,\cr
\label{secondd=1}
\eea
plug $d=0$ and the known leading order functions into (\ref{SSeqn}) to obtain
\bea
 -x_j^3 x_{j^3} \left(x_j 
\left[ {\partial f^{(1)}_{c,0}\over\partial x_{j^3}}
-{\partial f^{(1)}_{c,1}\over\partial x_{j^3}}
+{\partial f^{(1)}_{c,0}\over\partial x_j}
+{\partial f^{(1)}_{c,1}\over\partial x_j}\right]-2 f^{(1)}_{c,1}\right)\cr
 -x_j^5 
\left({\partial f^{(1)}_{c,0}\over\partial x_{j^3}}
+{\partial f^{(1)}_{c,1}\over\partial x_{j^3}}
+{\partial f^{(1)}_{c,0}\over\partial x_j}
-{\partial f^{(1)}_{c,1}\over\partial x_j}\right)
+\kappa_1 x_j x_{j^3}^2-\kappa_1 x_{j^3}^3=0\,,
\label{secondd=0}
\eea
and finally, plug $d=-2$ and the known leading order functions into (\ref{SSeqn}) to obtain
\bea
-x_j\left({\partial f^{(1)}_{c,-1}\over\partial x_{j^3}}
         -{\partial f^{(1)}_{c,-1}\over\partial x_j}\right) 
(x_j-x_{j^3})-2 f^{(1)}_{c,-1} x_{j^3}=0\, .
\label{secondd=-1}
\eea

We will now solve the above 6 partial differential equations simultaneously.
To start, sum (\ref{firstd=1}) and (\ref{secondd=1}) which leads to
\bea
  4\left(
x_j     {\partial f^{(1)}_{c,1}\over\partial x_j}+
x_{j^3} {\partial f^{(1)}_{c,1}\over\partial x_{j^3}}\right)
+2\kappa_1 {x_{j^3}^2\over x_j^3}=0\, .
\eea
The most general solution, regular at $x_{j^3}=0$ is
\bea
   f^{(1)}_{c,1}={\kappa_1\over 2}{x_{j^3}^2\over x_j^3}
                +\sum_{n=0}^\infty c_n {x_{j^3}^n\over x_j^n}\, .
\eea
Inserting this solution into (\ref{firstd=1}) we find
\bea
\sum_n 2c_n x_j^{3-n}x_{j^3}^{n-2}\left(nx_j^2-(n-2)x_{j^3}^2\right)=0\, .
\eea
Rearranging a little we find
\bea
\sum_{m=-2}^\infty 2c_{m+2}(m+2) x_j^{3-m}x_{j^3}^{m}
-\sum_{n=0}^\infty  2c_n (n-2)x_j^{3-n}x_{j^3}^n =0\, .
\nonumber
\eea
From the coefficient of $x_j x_{j^3}^2$ we have $4c_4=0$. 
From the coefficient of $x_j^{-1-2k} x_{j^3}^{4+2k}$ we have $(6+2k)c_{6+2k}=(4+2k)c_{4+2k}$ which together implies $c_{2k}=0$ for $k\ge 2$. 
From the coefficient of $x_j^3$ we have $2c_2=-2c_0$.
This just shifts the constant $\kappa_1$ appearing in $f^{(0)}_{c,1}$ by a term of $O({1\over \sqrt{m}})$ and we may as well set it to zero.
We shoud have expected this - as we described in the last section, this is one of the symmetries that are present in our equations.
By setting the coefficient of $x_j^4 x_{j^3}^{-1}$ to zero we find $c_1=0$ and from the coefficient of $x_j^{4-2k} x_{j^3}^{-1+2k}$ we find
$c_{2k+1}=0$ for $k>1$. 
Putting everything together we only get a solution if all the coefficients $c_n=0$.
Thus, we finally obtain
\bea
   f^{(1)}_{c,1}={\kappa_1\over 2}{x_{j^3}^2\over x_j^3}
\eea
which is indeed the correct answer.

Now, consider (\ref{firstd=-1}) and (\ref{secondd=-1}). From these two equations we can solve for 
${\partial f^{(1)}_{c,-1}\over\partial x_j}$ and for ${\partial f^{(1)}_{c,-1}\over\partial x_j}$
\bea
{\partial f^{(1)}_{c,-1}\over\partial x_j}=-{4 x_{j^3}^2 f^{(1)}_{c,-1} \over x_j (x_j^2-x_{j^3}^2)}\,,
\qquad
{\partial f^{(1)}_{c,-1}\over\partial x_{j^3}}={4 x_{j^3} f^{(1)}_{c,-1} \over x_j^2-x_{j^3}^2}\,.
\eea
These two equations are integrable - they give the same answer for ${\partial^2 f^{(1)}_{c,1}\over\partial x_j\partial x_{j^3}}$.
The only solution again corresponds to shifting $\kappa_1$, so that up to symmetry the most general solution is
\bea
   f^{(1)}_{c,-1}=0
\eea
which is again the correct answer.

Finally, consider (\ref{firstd=0}) and (\ref{secondd=0}). After plugging in the solution we found for
$f^{(1)}_{c,1}$ we find
\bea
2 x_j^4 \left(
{\partial f^{(1)}_{c,0}\over\partial x_{j^3}}-
{\partial f^{(1)}_{c,0}\over\partial x_j}\right)+2 \kappa_1 x_j x_{j^3}+3 \kappa_1 x_{j^3}^2=0
\eea
and
\bea
2 x_j^4 \left(
{\partial f^{(1)}_{c,0}\over\partial x_{j^3}}+
{\partial f^{(1)}_{c,0}\over\partial x_j}\right)+2 \kappa_1 x_j x_{j^3}-3 \kappa_1 x_{j^3}^2=0\,.
\eea
It is trivial to obtain the unique solution up to symmetry
\bea
  f^{(1)}_{c,0} = -{\kappa_1\over 2}{x_{j^3}^2\over x_j^3}
\eea
which is again correct. 
This reproduces the complete leading correction at one loop.

We can now check if our solution is hermittian which implies the following two conditions
\bea
    f^{(0)}_{c,a}(x_j,x_{j^3})= f^{(0)}_{c,-a}(x_j,x_{j^3})
\eea
and
\bea
    f^{(1)}_{c,a}(x_j,x_{j^3})= f^{(1)}_{c,-a}(x_j,x_{j^3})+a{\partial f^{(0)}_{c,-a}\over\partial x_j}\, .
\eea
Recall that at one loop we have
\bea
   f^{(0)}_{c,\pm 1}={\kappa_1\over 4}\left(1-{x_{j^3}^2\over x_j^2}\right),\qquad
  f^{(1)}_{c,1}={\kappa_1\over 2}{x_{j^3}^2\over x_j^3},\qquad f^{(1)}_{c,-1}=0\, .
\eea
It is a non-trivial fact that
\bea
    f^{(1)}_{c,1}(x_j,x_{j^3})= f^{(1)}_{c,-1}(x_j,x_{j^3})+{\partial f^{(0)}_{c,-1}\over\partial x_j}
\eea
so that our one loop solution is indeed Hermittian.

Finally, the next order is determined by the requirement that the $O(1)$ piece of (\ref{forfirst}) vanishes. 
Plugging in the solutions for $f^{(0)},f^{(1)}$ as well as $d=1$, we find
\bea
 16 x_j^6 \left({\partial f^{(2)}_{c,1}\over\partial x_{j^3}}
          -{\partial f^{(2)}_{c,1}\over\partial x_j}+x_{j^3}\kappa_1\right)
   +16 x_j^5 x_{j^3}
   \left(-{\partial f^{(2)}_{c,1}\over\partial x_{j^3}}
    +{\partial f^{(2)}_{c,1}\over\partial x_j}+x_{j^3}\kappa_1\right)\cr
  -16 x_j^4 x_{j^3} \left(-2 f^{(2)}_{c,1}
  +x_{j^3}^2\kappa_1\right)-16 x_j^7\kappa_1-x_j^3\kappa_1+25 x_j^2 x_{j^3}\kappa_1+25 x_j x_{j^3}^2\kappa_1-25 x_{j^3}^3\kappa_1=0\cr
  \label{fstdis1}
\eea
and
\bea
 -16 x_j^6 \left({\partial f^{(2)}_{c,1}\over\partial x_{j^3}}
           +{\partial f^{(2)}_{c,1}\over\partial x_j}
          +x_{j^3}\kappa_1\right)-16 x_j^5 x_{j^3}
   \left({\partial f^{(2)}_{c,1}\over\partial x_{j^3}}
   +{\partial f^{(2)}_{c,1}\over\partial x_j}-x_{j^3}\kappa_1\right)\cr
  +16 x_j^4 x_{j^3} \left(-2 f^{(2)}_{c,1}+x_{j^3}^2\kappa_1\right)-16
   x_j^7\kappa_1-x_j^3\kappa_1-25 x_j^2 x_{j^3}\kappa_1+25 x_j x_{j^3}^2\kappa_1+25 x_{j^3}^3\kappa_1=0\,.\cr
   \label{scndis1}
\eea
Summing (\ref{fstdis1}) and (\ref{scndis1}) we find
\bea
   x_j{\partial f^{(2)}_{c,1}\over\partial x_j}
+ x_{j^3} {\partial f^{(2)}_{c,1}\over\partial x_{j^3}}
- x_{j^3}^2\kappa_1 + x_j^2\kappa_1 +{\kappa_1\over 16x_j^2}-{25 x_{j^3}^2\over 16x_j^4}\kappa_1=0\,.
\eea
The general solution to this equation is (again we have required that the solution is regular at $x_{j^3}=0$)
\bea
   f^{(2)}_{c,1}={x_{j^3}^2\over 2}\kappa_1-{x_j^2\over 2}\kappa_1+{\kappa_1\over 32 x_j^2}-{25 x_{j^3}^2\over 32x_j^4}\kappa_1
           +\sum_{n=0}a_n {x_{j^3}^n\over x_j^n}\,.
\eea
Plugging this into (\ref{fstdis1}) we find
\bea
 \sum_{n=0}^\infty a_n  x_j^{-n-3} x_{j^3}^{n-1} \left(n x_j^2-(n-2) x_{j^3}^2\right)=0\,.
\eea
The most general solution to this equation is $a_0=-a_2$ and $a_n=0$ for $n\ne 0,2$.
You reach precisely the same conclusion if you use (\ref{scndis1}) instead of (\ref{fstdis1}).
Thus, our solution is
\bea
   f^{(2)}_{c,1}={x_{j^3}^2\over 2}\kappa_1-{x_j^2\over 2}\kappa_1+{\kappa_1\over 32 x_j^2}-{25 x_{j^3}^2\over 32x_j^4}\kappa_1
           +k_0 - k_0 {x_{j^3}^2\over x_j^2}\,.
\eea
Setting $k_0={1\over 2}$ and $\kappa_1=1$ we recover the answer from expanding the known dilatation operator coefficients.

Plugging in the solutions for $f^{(0)},f^{(1)}$ as well as $d=0$ we find
\bea
 16 x_j^6 \left({\partial f^{(2)}_{c,0}\over\partial x_{j^3}}
          +{\partial f^{(2)}_{c,1}\over\partial x_{j^3}}
          -{\partial f^{(2)}_{c,0}\over\partial x_j}
          +{\partial f^{(2)}_{c,1}\over\partial x_j}-x_{j^3}\kappa_1\right)\cr
     -16 x_j^5 x_{j^3}
   \left({\partial f^{(2)}_{c,0}\over\partial x_{j^3}}
   -{\partial f^{(2)}_{c,1}\over\partial x_{j^3}}
   -{\partial f^{(2)}_{c,0}\over\partial x_j}
   -{\partial f^{(2)}_{c,1}\over\partial x_j}+x_{j^3}\kappa_1\right)\cr
   +16 x_j^4 x_{j^3} \left(x_{j^3}^2\kappa_1-2 f^{(2)}_{c,1}\right)+16
   x_j^7\kappa_1+x_j^3\kappa_1+11 x_j^2 x_{j^3}\kappa_1-41 x_j x_{j^3}^2\kappa_1-43 x_{j^3}^3\kappa_1=0\cr
   \label{fstdis0}
\eea
and
\bea
 -16 x_j^6 \left({\partial f^{(2)}_{c,0}\over\partial x_{j^3}}
           +{\partial f^{(2)}_{c,1}\over\partial x_{j^3}}
           +{\partial f^{(2)}_{c,0}\over\partial x_j}
           -{\partial f^{(2)}_{c,1}\over\partial x_j}-x_{j^3}\kappa_1\right)\cr
    -16 x_j^5 x_{j^3}
   \left({\partial f^{(2)}_{c,0}\over\partial x_{j^3}}
   -{\partial f^{(2)}_{c,1}\over\partial x_{j^3}}
   +{\partial f^{(2)}_{c,0}\over\partial x_j}
   +{\partial f^{(2)}_{c,1}\over\partial x_j}+x_{j^3}\kappa_1\right)\cr
-16 x_j^4 x_{j^3} \left(x_{j^3}^2\kappa_1-2 f^{(2)}_{c,1}\right)+16
   x_j^7\kappa_1+x_j^3\kappa_1-11 x_j^2 x_{j^3}\kappa_1-41 x_j x_{j^3}^2\kappa_1+43 x_{j^3}^3\kappa_1=0\,.\cr
   \label{scddis0}
\eea
Now, summing (\ref{fstdis0}) and (\ref{scddis0})  we find
\bea
   -x_j {\partial f^{(2)}_{c,0}\over\partial x_j}
   -x_{j^3} {\partial f^{(2)}_{c,0}\over\partial x_{j^3}}
   +x_j {\partial f^{(2)}_{c,1}\over\partial x_j}
   +x_{j^3} {\partial f^{(2)}_{c,1}\over\partial x_{j^3}}
   -x_{j^3}^2\kappa_1+x_j^2\kappa_1+{1\over 16x_j^2}\kappa_1-{41 x_{j^3}^2\over 16x_j^4}\kappa_1=0\,.\cr
\eea
Plugging in the solution for $f^{(2)}_{c,1}$ that we constructed above, we find
\bea
    x_j {\partial f^{(2)}_{c,0}\over\partial x_j}
   +x_{j^3} {\partial f^{(2)}_{c,0}\over\partial x_{j^3}}
   +{x_{j^3}^2\over x_j^4}\kappa_1=0
\eea
which has the general solution
\bea
   f^{(2)}_{c,0}={x_{j^3}^2\over 2x_j^4}\kappa_1+\sum_{n=0}a_n {x_{j^3}^n\over x_j^n}\,.
\eea
Inserting this solution into (\ref{fstdis0}) we finally find
\bea
   f^{(2)}_{c,0}={x_{j^3}^2\over 2x_j^4}\kappa_1+2k_0 {x_{j^3}^2\over x_j^2}
\eea
where $k_0$ is the same constant that appeared above.

Finally, plugging in the solutions for $f^{(0)},f^{(1)}$ as well as $d=-2$ we find
\bea
 16 x_j^6 \left({\partial f^{(2)}_{c,-1}\over\partial x_{j^3}}
          +{\partial f^{(2)}_{c,-1}\over\partial x_j}+x_{j^3}\kappa_1\right)-16 x_j^5 x_{j^3}
   \left(-{\partial f^{(2)}_{c,-1}\over\partial x_{j^3}}
    -{\partial f^{(2)}_{c,-1}\over\partial x_j}+x_{j^3}\kappa_1\right)\cr
  -16 x_j^4 x_{j^3} \left(-2 f^{(2)}_{c,-1}+x_{j^3}^2\kappa_1\right)+16
   x_j^7\kappa_1+x_j^3\kappa_1+x_j^2 x_{j^3}\kappa_1-x_j x_{j^3}^2\kappa_1-x_{j^3}^3\kappa_1=0\cr
   \label{fstdism2}
\eea
and
\bea
 -16 x_j^6 \left({\partial f^{(2)}_{c,-1}\over\partial x_{j^3}}
           -{\partial f^{(2)}_{c,-1}\over\partial x_j}+x_{j^3}\kappa_1\right)-16 x_j^5 x_{j^3}
   \left(-{\partial f^{(2)}_{c,-1}\over\partial x_{j^3}}
    +{\partial f^{(2)}_{c,-1}\over\partial x_j}+x_{j^3}\kappa_1\right)\cr
    +16 x_j^4 x_{j^3} \left(-2 f^{(2)}_{c,-1}+x_{j^3}^2\kappa_1\right)+16
   x_j^7\kappa_1+x_j^3\kappa_1-x_j^2 x_{j^3}\kappa_1-x_j x_{j^3}^2\kappa_1+x_{j^3}^3\kappa_1=0\,.\cr
   \label{scndism2}
\eea
Summing (\ref{fstdism2}) and (\ref{scndism2}) we find
\bea
   x_j{\partial f^{(2)}_{c,-1}\over\partial x_j}
+ x_{j^3} {\partial f^{(2)}_{c,-1}\over\partial x_{j^3}}
- x_{j^3}^2\kappa_1 + x_j^2\kappa_1 +{\kappa_1\over 16x_j^2}-{x_{j^3}^2\over 16x_j^4}\kappa_1 =0\,.
\eea
The general solution to this equation is (again we have required that the solution is regular at $x_{j^3}=0$)
\bea
   f^{(2)}_{c,-1}={x_{j^3}^2\over 2}\kappa_1-{x_j^2\over 2}\kappa_1+{\kappa_1\over 32 x_j^2}-{x_{j^3}^2\over 32x_j^4}\kappa_1
           +\sum_{n=0}a_n {x_{j^3}^n\over x_j^n}\,.
\eea
Plugging this into (\ref{fstdism2}) we find
\bea
 \sum_{n=0}^\infty a_n  x_j^{-n-3} x_{j^3}^{n-1} \left(n x_j^2-(n-2) x_{j^3}^2\right)=0\,.
\eea
This is the equation we obtained above; the most general solution is $a_0=-a_2$ and $a_n=0$ for $n\ne 0,2$.
Thus, our solution is
\bea
   f^{(2)}_{c,-1}={x_{j^3}^2\over 2}\kappa_1-{x_j^2\over 2}\kappa_1+{\kappa_1\over 32 x_j^2}-{x_{j^3}^2\over 32x_j^4}\kappa_1
           +\tilde{k}_0 - \tilde{k}_0 {x_{j^3}^2\over x_j^2}\,.
\eea
Setting $\tilde{k}_0={1\over 2}$ we recover the answer from expanding the known dilatation operator coefficients.

If we now study the $d=-1$ equation we can prove that $k_0=\tilde{k}_0$. Thus, in summary we have
\bea
   f^{(2)}_{c,1}={x_{j^3}^2\over 2}\kappa_1-{x_j^2\over 2}\kappa_1+{\kappa_1\over 32 x_j^2}-{25 x_{j^3}^2\over 32x_j^4}\kappa_1
           +k_0 - k_0 {x_{j^3}^2\over x_j^2}\,,
\eea
\bea
   f^{(2)}_{c,0}={x_{j^3}^2\over 2x_j^4}\kappa_1+2k_0 {x_{j^3}^2\over x_j^2}\,,
\eea
\bea
   f^{(2)}_{c,-1}={x_{j^3}^2\over 2}\kappa_1-{x_j^2\over 2}\kappa_1+{1\over 32 x_j^2}\kappa_1-{x_{j^3}^2\over 32x_j^4}\kappa_1
           +k_0 - k_0 {x_{j^3}^2\over x_j^2}\,.
\eea

Collecting the results we have found above, we have the three functions above as well as
\bea
   f^{(0)}_{c,1}  =  {\kappa_1\over 4}  \left( 1-{x_{j^3}^2\over x_j^2}\right),\qquad
   f^{(0)}_{c,0}  =-{\kappa_1\over 2} \left( 1-{x_{j^3}^2\over x_j^2}\right),\qquad
   f^{(0)}_{c,-1}  =  {\kappa_1\over 4}  \left( 1-{x_{j^3}^2\over x_j^2}\right)
\eea
\bea
   f^{(1)}_{c,1}={\kappa_1\over 2}{x_{j^3}^2\over x_j^3},\qquad
  f^{(1)}_{c,0} = -{\kappa_1\over 2}{x_{j^3}^2\over x_j^3},\qquad
  f^{(1)}_{c,-1} =0\, .
\eea
Requiring that the smallest eigenvalue of the one loop dilatation operator is zero determines $k_0=0$.
Thus, up to an overall normalization which our argument can't determine, we have again reproduced (\ref{basicdil}).

\subsection{General Discussion}

In this section we will extended our arguments to higher loops.
More specifically, in the language of the discussion towards the end of section 4, our goal is to construct the most general solution up to symmetry.
Recall that we have already determined (see (\ref{ploopone}) and (\ref{plooptwo}) above)
\bea
   f^{(0)}_{c,p}= \kappa_p \left( 1-{x_{j^3}^2\over x_j^2}\right)^p\,,
\eea
\bea
   f^{(0)}_{c,p-1} =-2p\kappa_p \left( 1-{x_{j^3}^2\over x_j^2}\right)^p + \kappa_{p-1}\left( 1-{x_{j^3}^2\over x_j^2}\right)^{p-1}\,.
\eea
Plug $d=p$ and the known leading order functions into (\ref{FFeqn}) to obtain
\bea
    x_j (x_j-x_{j^3}) \left(x_j \left[{\partial f^{(1)}_{c,p}\over\partial x_{j^3}}
  -{\partial f^{(1)}_{c,p}\over\partial x_j}\right] (x_j-x_{j^3})+2 f^{(1)}_{c,p} p x_{j^3}\right)\cr
  +\kappa_p \left(2 (p-1)
   x_j^4+x_{j^3}^2 \left(2 (p-1) x_j^2+p (p+1)\right)-2 x_j x_{j^3} \left(2 (p-1) x_j^2+p (p+1)\right)\right)
   \left(1-\frac{x_{j^3}^2}{x_j^2}\right)^p=0\,.\nonumber
\eea
Plug $d=p$ and the known leading order functions into (\ref{SSeqn}) to obtain
\bea
    \kappa_p \left(1-\frac{x_{j^3}^2}{x_j^2}\right)^p 
   \left(2 (p-1) x_j^4+4 (p-1) x_j^3 x_{j^3}+2 (p-1) x_j^2 x_{j^3}^2+2 p (p+1)
    x_j x_{j^3}+p (p+1) x_{j^3}^2\right)\cr
  -x_j (x_j+x_{j^3}) \left(x_j \left[{\partial f^{(1)}_{c,p}\over\partial x_{j^3}}
                          +{\partial f^{(1)}_{c,p}\over\partial x_j}\right] (x_j+x_{j^3})+2 f^{(1)}_{c,p}
   p x_{j^3}\right)=0\,.\nonumber
\eea
Summing these two we obtain
\bea
   x_j{\partial f^{(1)}_{c,p}\over\partial x_j}+x_{j^3}{\partial f^{(1)}_{c,p}\over\partial x_{j^3}}
-\kappa_p\left( 1-{x_{j^3}^2\over x_j^2}\right)^{p-1}\left[
2(p-1)x_j
-p(p+1){x_{j^3}^2\over x_j^3}
-2(p-1){x_{j^3}^2\over x_j}
\right]=0\,.\nonumber
\eea
The general solution to this equation, that is regular at $x_{j^3}=0$ is
\bea
   f^{(1)}_{c,p}=
\kappa_p\left( 1-{x_{j^3}^2\over x_j^2}\right)^{p-1}\left[
2(p-1)x_j
+p(p+1){x_{j^3}^2\over x_j^3}
-2(p-1){x_{j^3}^2\over x_j}
\right]
+\sum_{n=0}^\infty a_n {x_{j^3}^n\over x_j^n}\,.\nonumber
\eea
Plugging this back into the first of our equations we find
\bea
  \sum_{n=0}^\infty a_n x_j^{-n} x_{j^3}^{n} \left[n (x_j-x_{j^3}) (x_j+x_{j^3})+2 p x_{j^3}^2\right]=0\,.
\eea
The only solution is $a_n=0$ so that
\bea
   f^{(1)}_{c,p}=
\kappa_p\left( 1-{x_{j^3}^2\over x_j^2}\right)^{p-1}\left[
2(p-1)x_j
+p(p+1){x_{j^3}^2\over x_j^3}
-2(p-1){x_{j^3}^2\over x_j}
\right]\,.
\eea

Now, plug $d=p$ and the known leading order functions and $f^{(1)}_{c,p}$ into (\ref{FFeqn}) to obtain
\bea
 x_j (x_j-x_{j^3})^2 (x_j+x_{j^3}) \left[x_j \left({\partial f^{(1)}_{c,p-1}\over\partial x_{j^3}}
                                        -{\partial f^{(1)}_{c,p-1}\over\partial x_j}\right)(x_j-x_{j^3})
  +2 f^{(1)}_{c,p-1} (p-1) x_{j^3}\right]\cr
-\left(1-\frac{x_{j^3}^2}{x_j^2}\right)^p 
\left(x_j^2 x_{j^3}^2 \left(p (\kappa_{p} (6 p+3)+\kappa_{p-1} (-p)+\kappa_{p-1}+2 (p-3) p-3)
\right.\right.\cr
\left.\left.
-2 x_j^2 \left(\kappa_{p} \left(4 p^2-6 p+4\right)+\kappa_{p-1} p-2 (\kappa_{p-1}+p)\right)\right)
\right.\cr
+2 x_j^6 (2 \kappa_{p}-(p-2) (\kappa_{p-1}-2 p))+x_j^3
   x_{j^3} \left(4 x_j^2 ((\kappa_{p}-1) p (2 p-3)+\kappa_{p-1} (p-2))
\right.\cr
\left.
-p (4 \kappa_{p} p+\kappa_{p}-2 \kappa_{p-1} (p-1)+4 (p-1) p-1)\right)\cr
+p x_j
   x_{j^3}^3 \left(4 \kappa_{p} p^2-4 (\kappa_{p}-1) (2 p-3) x_j^2+4 \kappa_{p} p+\kappa_{p}+4 p^2-4 p-1\right)\cr
  -(\kappa_{p}-1) p x_j^4-2 x_{j^3}^4
   \left(p (\kappa_{p} (p+1) (2 p+1)+(p-3) p-1)\right.\cr
\left.\left.
-2 (p-1) x_j^2 (\kappa_{p} (2 p-1)-p)\right)\right)=0\cr
\eea
and plug $d=p$ and the known leading order functions and $f^{(1)}_{c,p}$ into (\ref{SSeqn}) to obtain
\bea
-x_j (x_j-x_{j^3}) (x_j+x_{j^3})^2 \left[ x_j
   \left({\partial f^{(1)}_{c,p-1}\over\partial x_{j^3}}
   +{\partial f^{(1)}_{c,p-1}\over\partial x_j}\right) 
  (x_j+x_{j^3})+2 f^{(1)}_{c,p-1} (p-1) x_{j^3}\right]\cr
 \left(1-\frac{x_{j^3}^2}{x_j^2}\right)^p \left(x_j^2 x_{j^3}^2 \left(2 x_j^2 \left[\kappa_{p} \left(4 p^2-6 p+4\right)
\right.\right.\right.\cr
\left.\left.
+\kappa_{p-1} p-2(\kappa_{p-1}+p)\right]-p\left[\kappa_{p}(6 p+3)+\kappa_{p-1}(-p)+\kappa_{p-1}+2 (p-3) p-3\right]\right)\cr
+2 x_j^6 ((p-2) (\kappa_{p-1}-2 p)-2 \kappa_{p})+x_j^3 x_{j^3}
   \left(4 x_j^2 ((\kappa_{p}-1) p (2 p-3)+\kappa_{p-1} (p-2))\right.\cr
\left.
-p (4 \kappa_{p} p+\kappa_{p}-2 \kappa_{p-1} (p-1)+4 (p-1) p-1)\right)+p x_j x_{j^3}^3 \left(4
   \kappa_{p} p^2\right.\cr
\left.
-4 (\kappa_{p}-1) (2 p-3) x_j^2+4 \kappa_{p} p+\kappa_{p}+4 p^2-4 p-1\right)+(\kappa_{p}-1) p x_j^4\cr
\left.+2 x_{j^3}^4 \left(p (\kappa_{p} (p+1) (2 p+1)+(p-3) p-1)-2 (p-1) x_j^2 (\kappa_{p} (2 p-1)-p)\right)\right)=0\,.\cr
\eea
Summing these two we obtain
\bea
 x_{j^3} {\partial f^{(1)}_{c,p-1}\over\partial x_{j^3}}
+x_j     {\partial f^{(1)}_{c,p-1}\over\partial x_j} 
+\left(1-\frac{x_{j^3}^2}{x_j^2}\right)^{p-2} F(x_j,x_{j^3})=0\,,
\eea
where
\bea
F(x_j,x_{j^3})=
\frac{x_{j^3}^2}{x_j} \left(-8 \kappa_p+2 \kappa_{p-1} p-4 \kappa_{p-1}-8 p^2\kappa_p +16 p\kappa_p\right)\cr
+x_j \left(4 \kappa_p-2 \kappa_{p-1} p+4\kappa_{p-1}+4 p^2\kappa_p-8 p\kappa_p\right)+2 p^3\kappa_p\frac{x_{j^3}^4}{x_j^5}\cr
+\frac{x_{j^3}^2}{x_j^3} \left(\kappa_{p-1} p^2-\kappa_{p-1} p-2 p^3\kappa_p\right)
+\left(4\kappa_p+4 p^2\kappa_p-8 p\kappa_p\right)\frac{x_{j^3}^4 }{x_j^3}\,.
\nonumber
\eea
The general solution to this equation, which is regular at $x_{j^3}=0$ is
\bea
f^{(1)}_{c,p-1}=\left(1-\frac{x_{j^3}^2}{x_j^2}\right)^{p-2} G(x_j,x_{j^3})+\sum_{n=0}^\infty a_n {x_{j^3}^n\over x_j^n}
\eea
where
\bea
G(x_j,x_{j^3})=
-\frac{x_{j^3}^2}{x_j}  \left(-8 \kappa_p+2 \kappa_{p-1} p-4 \kappa_{p-1}-8 p^2\kappa_p +16 p\kappa_p\right)\cr
-x_j \left(4 \kappa_p-2 \kappa_{p-1} p+4\kappa_{p-1}+4 p^2\kappa_p-8 p\kappa_p\right)
+2 p^3\kappa_p \frac{x_{j^3}^4}{x_j^5}\cr
+\frac{x_{j^3}^2}{x_j^3} \left(\kappa_{p-1} p^2-\kappa_{p-1} p-2 p^3\kappa_p\right)
-\left(4\kappa_p+4 p^2\kappa_p-8 p\kappa_p\right)\frac{x_{j^3}^4 }{x_j^3}\,.
\nonumber
\eea
Plugging this back into the first equation above we find
\bea
x_j^{-n-1} x_{j^3}^{n-1} \left[ a_n n x_j^2-a_n x_{j^3}^2 (n-2 p+2)\right]=0
\eea
which forces $a_n=0$.

Now, study the equation obtained by plugging $d=-p-1$ and the known leading order functions into (\ref{FFeqn}) to obtain
\bea
 x_j (x_j+x_{j^3}) \left(x_j \left[{\partial f^{(1)}_{c,-p}\over\partial x_{j^3}}
                        +{\partial f^{(1)}_{c,-p}\over\partial x_j}\right] (x_j+x_{j^3})+2 f^{(1)}_{c,-p} p x_{j^3}\right)\cr
  +\kappa_p (p-1) \left[x_{j^3}^2
   \left(p+2 x_j^2\right)+2 x_j x_{j^3} \left(p+2 x_j^2\right)+2 x_j^4\right] \left(1-\frac{x_{j^3}^2}{x_j^2}\right)^p=0\,.\nonumber
\eea
Plug $d=p$ and the known leading order functions into (\ref{SSeqn}) to obtain
\bea
  \kappa_p (p-1) \left(1-\frac{x_{j^3}^2}{x_j^2}\right)^p \left[x_{j^3}^2 \left(p+2 x_j^2\right)-2 x_j x_{j^3} \left(p+2
   x_j^2\right)+2 x_j^4\right]\cr
 -x_j (x_j-x_{j^3}) \left(x_j \left[ {\partial f^{(1)}_{c,-p}\over\partial x_{j^3}}
                         -{\partial f^{(1)}_{c,-p}\over\partial x_j}\right] (x_j-x_{j^3})+2 f^{(1)}_{c,-p} p
   x_{j^3}\right)=0\,.\nonumber
\eea
Summing these two we obtain
\bea
   x_j^3 \left({\partial f^{(1)}_{c,-p}\over\partial x_{j^3}} x_{j^3}
         +{\partial f^{(1)}_{c,-p}\over\partial x_j} x_j\right)-\kappa_p (p-1)
   \left(1-\frac{x_{j^3}^2}{x_j^2}\right)^{p-2} 
   \left[x_{j^3}^2 \left(p+2 x_j^2\right)-2 x_j^4\right]=0\,.\nonumber
\eea
The general solution to this equation, that is regular at $x_{j^3}=0$ is
\bea
   f^{(1)}_{c,-p}=\kappa_p (p-1)\left(1-{x_{j^3}^2\over x_j^2}\right)^{p-1}
                  \left(-p{x_{j^3}^2\over x_j^3}+{2x_{j^3}^2\over x_j}-2x_j\right)
                  +\sum_{n=0}^\infty a_n{x_{j^3}^n\over x_j^n}\,.
\eea
Plugging this back into the first equation above we learn that $a_n=0$.

The only results we need from the above analysis are
\bea
   f^{(0)}_{c,\pm p}= \kappa_p \left( 1-{x_{j^3}^2\over x_j^2}\right)^p\,,
\eea
\bea
   f^{(1)}_{c,p}=
\kappa_p\left( 1-{x_{j^3}^2\over x_j^2}\right)^{p-1}\left[
2(p-1)x_j
+p(p+1){x_{j^3}^2\over x_j^3}
-2(p-1){x_{j^3}^2\over x_j}
\right]\,,
\eea
\bea
   f^{(1)}_{c,-p}=\kappa_p (p-1)\left(1-{x_{j^3}^2\over x_j^2}\right)^{p-1}
                  \left(-p{x_{j^3}^2\over x_j^3}+{2x_{j^3}^2\over x_j}-2x_j\right)+\sum_{n=0}^\infty a_n{x_{j^3}^n\over x_j^n}\,.
\eea
Now, computing
\bea
   f^{(1)}_{c,p}(x_j,x_{j^3})-f^{(1)}_{c,-p}(x_j,x_{j^3})-p{\partial f^{(0)}_{c,-p}\over\partial x_j}=4\kappa_p (p-1)x_j
                       \left(1-{x_{j^3}^2\over x_j^2}\right)^p
\eea
we see that the only time that we get a Hermittian solution is when $p=1$.
Thus we are forced to set $f^{(0)}_{c,\pm p}=f^{(1)}_{c,\pm p}=0$ which then implies that $f^{(2)}_{c,\pm p}=0$.
We now apply the same argument to conclude that $f^{(0)}_{c,\pm p\mp 1}=f^{(1)}_{c,\pm p\mp 1}=f^{(2)}_{c,\pm p\mp 1}=0$ and keep going. 
Finally, when we get to $f^{(0)}_{c,\pm 1},f^{(1)}_{c,\pm 1},f^{(2)}_{c,\pm 1}$, we will find the one loop answer. 
This proves that the form of the piece of the dilatation operator that acts on the $Y$ fields is not corrected at any higher loop order.

{\vskip 1.0cm}

\noindent
{\it Acknowledgements:} 
This work is supported by the South African Research Chairs
Initiative of the Department of Science and Technology and the National Research Foundation.
Any opinion, findings and conclusions or recommendations expressed in this material
are those of the authors and therefore the NRF and DST do not accept any liability
with regard thereto.
RdMK thanks Perimeter Institute for hospitality and for providing an excellent research environment.
Research at Perimeter Institute is supported by the Government of Canada
through Industry Canada and by the Province of Ontario through the Ministry of Economic Development \&
Innovation.

\section{Appendix: The relation between $f^{(n,m)}_{c,d}(x_j,x_{j^3})$ and $f^{(n-1,m+1)}_{c,d}(x_j,x_{j^3})$}
\label{subtleshift}

In this appendix we derive a relation between $f^{(n,m)}_{c,d}(x_j,x_{j^3})$ and $f^{(n-1,m+1)}_{c,d}(x_j,x_{j^3})$ that is used
extensively in section \ref{higherloopanalysis}.
To make the discussion concrete we will study $f^{(n,m)}_{c,0}(x_j,x_{j^3})$ which is the continuum limit function corresponding to the 
following dilatation operator matrix element
\bea
   -{1\over 2}\left[ m -{(m+2)(j^3)^2\over j(j+1)}\right]
   \label{diagmatelem}
\eea
This becomes the following function
\bea
  f^{(n,m)}_{c,0}(x_j,x_{j^3})=   -{1\over 2}\left[ m -{(m+2)(\sqrt{m}x_{j^3})^2\over \sqrt{m}x_j(\sqrt{m}x_j+1)}\right]
\eea
We have the series expansion
\bea
   f^{(n,m)}_{c,0}(x_j,x_{j^3})=\sum_{q=0}^\infty m^{1-{q\over 2}}f^{(m)}_{c,0}(x_j,x_{j^3})
\eea

When we replace $m\to m+1$, we do so {\it without} changing $j$ and $j^3$ - it is the expression (\ref{diagmatelem}) with
$m\to m+1$ that solves the correct recursion relation. 
We must use the same definition of $x_j$ and $x_{j^3}$ for both $f^{(n,m)}_{c,0}(x_j,x_{j^3})$ and $  f^{(n-1,m+1)}_{c,0}(x_j,x_{j^3})$, which
implies that the new dilatation operator matrix element
\bea
   -{1\over 2}\left[ m+1 -{(m+1+2)(j^3)^2\over j(j+1)}\right]
\eea
leads to the following function
\bea
  f^{(n-1,m+1)}_{c,0}(x_j,x_{j^3})=   -{1\over 2}\left[ m+1 -{(m+1+2)(\sqrt{m}x_{j^3})^2\over \sqrt{m}x_j(\sqrt{m}x_j+1)}\right]\, .
\eea
We can get this function from $f^{(n,m)}_{c,0}(x_j,x_{j^3})$
by (i) shifting every $m\to m+1$ and then (ii) rescaling $x_j\to\sqrt{m\over m+1}x_j$ and
$x_{j^3}\to\sqrt{m\over m+1}x_{j^3}$. In summary
\bea
  f^{(n,m)}_{c,d}(x_j,x_{j^3})=\sum_{q=0}^\infty m^{1-{q\over 2}}f^{(m)}_{c,d}(x_j,x_{j^3})\, ,\cr
  f^{(n-1,m+1)}_{c,d}(x_j,x_{j^3})=\sum_{q=0}^\infty (m+1)^{1-{q\over 2}}f^{(m)}_{c,d}(\sqrt{m\over m+1}x_j,\sqrt{m\over m+1}x_{j^3})\, .
\eea
Finally, note that
\bea
   \sqrt{m\over m+1}=1-{1\over 2m}+{3\over 8m^2}+...
\eea

\end{document}